%% file: main.tex
\documentclass[compsoc,conference,a4paper,10pt,times]{IEEEtran}
\IEEEoverridecommandlockouts
\usepackage{cite}
\usepackage{amsmath,amssymb,amsfonts}
\usepackage{algorithmic}
\usepackage{graphicx}
\usepackage{textcomp}
\usepackage{bmpsize}
\usepackage{lipsum}
\usepackage[colorlinks=true,urlcolor=black]{hyperref}
\def\BibTeX{{\rm B\kern-.05em{\sc i\kern-.025em b}\kern-.08em
    T\kern-.1667em\lower.7ex\hbox{E}\kern-.125emX}}

\usepackage{subcaption}
\usepackage{url}
\usepackage{hyperref}
\hypersetup{                    
  colorlinks,
  linkcolor={black!80!black},
  citecolor={red!70!black},
  urlcolor={blue!70!black}
}
\usepackage{enumitem}
\usepackage{tcolorbox}
\tcbuselibrary{listings,breakable}
\usepackage[font=small]{caption}
\usepackage{booktabs}
\usepackage{colortbl}
\usepackage{multirow}
\usepackage{array}
\usepackage{tabularx}
\usepackage{cleveref}
\usepackage{placeins}
\usepackage{float}
\usepackage{textcomp}
\usepackage{xspace}
\usepackage{xcolor}
\usepackage{caption}
\usepackage{url}
\usepackage{xurl}
\usepackage{microtype}
\usepackage{svg}
\usepackage{wasysym}

\begin{document}

\title{UEFI Memory Forensics:\\A Framework for UEFI Threat Analysis}

\author{\IEEEauthorblockN{1\textsuperscript{st} Kalanit Suzan Segal$^*$}
\IEEEauthorblockA{\textit{Ben Gurion University of the Negev} \\
Be'er Sheva, Israel \\
kalanits@post.bgu.ac.il}
\and
\IEEEauthorblockN{2\textsuperscript{nd} Hadar Cochavi Gorelik$^*$}
\IEEEauthorblockA{\textit{Ben Gurion University of the Negev} \\
Be'er Sheva, Israel \\
hadarcoc@post.bgu.ac.il}
\and
\IEEEauthorblockN{3\textsuperscript{rd} Oleg Brodt}
\IEEEauthorblockA{\textit{Ben Gurion University of the Negev} \\
Be'er Sheva, Israel \\
bolegb@bgu.ac.il}
\and
\IEEEauthorblockN{4\textsuperscript{th} Yuval Elbahar}
\IEEEauthorblockA{\textit{Ben Gurion University of the Negev} \\
Be'er Sheva, Israel \\
yuvalelb@post.bgu.ac.il}
\and
\IEEEauthorblockN{5\textsuperscript{th} Yuval Elovici}
\IEEEauthorblockA{\textit{Ben Gurion University of the Negev} \\
Be'er Sheva, Israel \\
elovici@bgu.ac.il}
\and
\IEEEauthorblockN{6\textsuperscript{th} Asaf Shabtai}
\IEEEauthorblockA{\textit{Ben Gurion University of the Negev} \\
Be'er Sheva, Israel \\
shabtaia@bgu.ac.il}
}

\maketitle
\renewcommand{\thefootnote}{\fnsymbol{footnote}}
\footnotetext[1]{These authors contributed equally to this work.}
\renewcommand{\thefootnote}{\arabic{footnote}}

\maketitle

\begin{abstract}
\input{Sections/Abstract}
\end{abstract}


\input{Sections/Introduction}
\input{Sections/Background}
\input{Sections/ThreatModel}
\input{Sections/Methodology}
\input{Sections/Evaluation}
\input{Sections/RelatedWork}
\input{Sections/Limitations}
\input{Sections/Conclusions}
\input{Sections/Acknowledgment}

\bibliographystyle{IEEEtran}
\bibliography{references}

\appendices
\input{Sections/Appendix}

\end{document}

%% file: Sections/Abstract.tex
Modern computing systems rely on the Unified Extensible Firmware Interface (UEFI), which has replaced the legacy Basic Input/Output System (BIOS) as the firmware standard for the modern boot process. Although the UEFI represents a significant advancement in system firmware, it is increasingly targeted by threat actors seeking to exploit its execution environment and take advantage of its persistence mechanisms. While some security-related analysis of UEFI components has been performed--primarily via debugging and runtime behavior testing--to the best of our knowledge, no prior study has specifically addressed the capturing and analysis of volatile UEFI runtime memory to detect malicious exploitation during the pre-OS phase.
This gap in UEFI forensic tools limits the ability to conduct in-depth security analysis in pre-OS environments. Such a gap is particularly surprising, given that memory forensics is widely regarded as foundational to modern incident response, as reflected by the popularity of above-OS memory analysis frameworks, such as Rekall, Volatility, and MemProcFS. 
To address the lack of below-OS memory forensics, we introduce a framework for UEFI memory forensics. The proposed framework consists of two components: \textit{UEFIMemDump}, a memory acquisition tool, and \textit{UEFIDumpAnalysis}, an extendable collection of analysis modules capable of detecting malicious activities such as function pointer hooking, inline hooking, malicious image loading, and gadget-based control-flow manipulation.
Our proof-of-concept implementation demonstrates the framework's ability to detect modern UEFI threats, such as Thunderstrike, CosmicStrand, and Glupteba bootkits. By providing an open-source solution, our work enables researchers and practitioners to investigate firmware-level threats, develop additional analysis modules, and advance overall below-OS security through UEFI memory analysis.

%% file: Sections/Introduction.tex
\section{Introduction}
\label{sec:introduction}
The Unified Extensible Firmware Interface (UEFI) \cite{UEFI2024} has replaced the legacy Basic Input/Output System (BIOS) as the standard for firmware, addressing the increasing demands of modern computing systems \cite{UEFI2023}. This transition was driven by fundamental limitations in the BIOS architecture, including its 16-bit operational mode, 1MB addressable memory constraint, and lack of modularity. Since its introduction in the early 2000s, UEFI has evolved significantly, providing a modular and extensible architecture that bridges the gap between hardware initialization and operating systems. Its adoption spans various platforms, from low-cost Raspberry Pi devices to mainstream laptops and desktops, and extends to high-performance enterprise servers, reflecting its ubiquity in contemporary computing environments \cite{sentinelone_uefi_dumping}.

UEFI’s central role in the architecture of modern computing makes it critical for security. It resides in the most fundamental level of the security stack, operating directly above the hardware, making it an attractive target for threat actors. Its high-privilege execution environment, persistence across reboots, and control over the boot process multiply the potential impact of a compromise at this layer, which would enable attackers to bypass kernel-level and hypervisor-based defenses, thereby threatening the security of the entire computer system.

To improve the security of the UEFI, the UEFI Forum \cite{UEFIForum}, along with industry and the research community \cite{TPM2, eclypsium_secure_boot_dbx, CapsuleUpdate}, has established security specifications and mechanisms for the modern boot sequence. UEFI-compliant firmware must incorporate several layered security mechanisms, beginning with UEFI Secure Boot \cite{eclypsium_secure_boot_dbx}, which prevents unauthorized boot components from executing by verifying digital signatures against a database of trusted certificates. This protection counters boot-time attacks that load malicious bootloaders, firmware-level drivers, and other compromised components. UEFI Capsule Updates \cite{CapsuleUpdate} provide an additional security layer through cryptographically signed and validated firmware updates, ensuring update integrity and authenticity. In addition, hardware-backed protections through Trusted Platform Modules (TPMs) \cite{TPM2} establish a root of trust, enabling secure key storage and attestation capabilities throughout the boot sequence. 
Together, these mechanisms form a Trusted Computing Base (TCB) \cite{TCB} designed to protect the firmware layer from malicious activities.

Despite the various security mechanisms in place, UEFI security measures' effectiveness depends heavily on proper implementation and configuration. In practice, Secure Boot is frequently disabled or misconfigured, either to support legacy software or due to users' lack of security awareness. Even when enabled, attackers can exploit its predefined policies and trusted keys, as demonstrated by the BlackLotus bootkit, which bypassed Secure Boot by leveraging compromised signed bootloaders \cite{blacklotus}. Similarly, UEFI Capsule Updates, despite their cryptographic protections, remain vulnerable to supply chain attacks, as demonstrated by the ShadowHammer campaign, in which attackers distributed malicious firmware updates through a compromised update process  \cite{shadowhammer}.

UEFI-based threats have evolved beyond implementation vulnerabilities, targeting the runtime environment directly. MoonBounce demonstrated persistence by residing in Serial Peripheral Interface (SPI) flash memory and redirecting execution flow during runtime to hook boot services \cite{MoonBounce_2022}. 
CosmicStrand advanced these techniques by altering runtime service structures to inject malware into the operating system \cite{CosmicStrand_2022}, while Glupteba further extended the approach by patching multiple boot components to disable security controls \cite{Glupteba_2024}.

In this context, aside from Secure Boot's signature verification, no security controls enable runtime analysis during the boot process, allowing malicious code to run without restrictions or security monitoring during the UEFI phase, as long as it passed or bypassed simple signature verification.

The detection of such runtime-based attacks is particularly challenging due to UEFI memory's dynamic nature. While runtime services persist after boot, many critical memory allocations from the Driver Execution Environment (DXE) phase are deallocated at OS initialization~\cite{zimmer2017beyond}. Traditional post-boot security tools cannot analyze these transient memory regions, creating a blind spot for defenders. 

These security gaps are particularly concerning, as memory forensics, arguably a linchpin of digital forensics and incident response, lacks dedicated tools for UEFI memory analysis. 
While existing memory analysis tools~\cite{volatility_foundation,rekall,lime,ftk_imager,memoryze, memprocfs_2023} are effectively used to detect various OS-level threats, they cannot address the pre-boot phase where the UEFI operates.

To address this gap, this paper introduces a novel framework for UEFI memory forensics, enabling the analysis of UEFI memory during the pre-boot phase. Our approach combines memory acquisition and analysis to detect malicious modifications in UEFI structures. Specifically, the contributions of this work are:

\noindent\textbf{UEFI Memory Analysis Framework}: We present the first memory analysis framework dedicated to the UEFI, enabling detection of unauthorized modifications in runtime control structures that evade traditional security measures. The framework consists of two main components:
    \begin{itemize} [topsep=0pt,noitemsep,leftmargin=*]
        \item \textbf{UEFI Memory Capture Module}: A specialized memory acquisition capability implemented as both a DXE driver and a UEFI shell application, capturing complete system memory snapshots during UEFI execution before OS initialization.
        \item \textbf{UEFI Memory Analysis Module}: An extendable suite of analysis modules operating on captured memory dumps to enable practical UEFI memory forensics, including:
        \begin{itemize}[topsep=0pt,noitemsep,leftmargin=*]
            \item \textbf{Function Pointer Hooking Detection Module} for identifying unauthorized modifications to service tables' function pointers;
            \item \textbf{Inline Hooking Detection Module} for discovering code-level execution redirections;
            \item \textbf{UEFI Image Carving Module} for extracting and analyzing UEFI images and boot-level executables;
            \item \textbf{Gadget-Based Control-Flow Detection Module} for identifying suspicious sequences of values that correspond to chained control-flow targets inside DXE drivers.
        \end{itemize}
    \end{itemize}
\noindent\textbf{Open-Source Implementation}: We make the framework's source code publicly available to the security community with the aim of encouraging security analysts to adopt it in their investigations and security researchers to expand it further with additional analysis modules.\footnote{\url{https://github.com/UefiMemAnalysis/UefiMemAnalysis}}\\

\noindent We evaluate our framework using attack techniques employed by modern bootkits such as Glupteba \cite{Glupteba_2024}, MoonBounce \cite{MoonBounce_2022}, and CosmicStrand \cite{CosmicStrand_2022}, as well as proof-of-concept (PoC) exploits like EFIGuard \cite{EfiGuard} and ThunderStrike \cite{Thunderstrike}, within a dedicated testbed. Our comprehensive evaluation demonstrates the framework's effectiveness in detecting UEFI threats that leverage hooking, malicious image loading, and gadget-based control-flow manipulation, while also showing that false positives arise across the detection modules. Each module therefore incorporates configurable filtering mechanisms, including GUID-based whitelists for hooking detection and tunable threshold parameters for gadget-based detection, to reduce them to manageable levels.

%% file: Sections/Background.tex
\section{Background}
\label{sec:background}

\subsection{Memory Forensics}
Memory forensics focuses on analyzing volatile RAM to uncover evidence of malicious activity and extract artifacts. It has gained prominence due to advanced threats such as fileless malware and rootkits that operate entirely in memory, evading traditional endpoint security tools.

In the OS context, memory forensics involves two stages: collection using tools such as FTK Imager~\cite{ftk_imager} or LiME~\cite{lime}, followed by analysis using frameworks like Volatility~\cite{volatility_foundation} and Rekall~\cite{rekall}. These frameworks enable examination of running processes, detection of hooks, and reconstruction of events. However, existing memory analysis frameworks focus on OS-level analysis and are designed to parse OS-specific data structures.

\subsubsection{The Gap in UEFI Memory Forensics}
Memory forensics in the context of UEFI firmware presents unique challenges due to its distinct execution environment. Unlike OS-level memory forensics, which benefits from an established knowledge-base, tools and frameworks, analyzing memory during the UEFI phase requires dedicated techniques and tools that understand UEFI-specific data structures and memory organization. While some tools like RDFU \cite{vuksan2013press} attempt to detect UEFI threats through runtime scanning, comprehensive memory forensics capabilities for the UEFI remain underdeveloped. This gap is particularly concerning, as forensic memory analysis provides deeper visibility into the system state and potential threats than runtime scanning approaches.

\subsubsection{Memory Acquisition Evasion by Malware}
Memory acquisition serves as the foundation of forensic investigation, capturing a snapshot of system memory for subsequent analysis. However, the trustworthiness of a memory dump depends on both the acquisition method and the extent of the attacker's control at the time of capture. 
In general, memory acquisition techniques can be divided into the following categories:

\noindent\textbf{Physical-Based Acquisition.}
Physical acquisition includes cold-boot and hardware-based DMA methods.
Cold-boot techniques exploit DRAM remanence, where memory contents persist briefly after power loss, allowing recovery without OS interaction~\cite{halderman2008lest}. 
However, this approach requires precise timing and physical access, and modern DRAM technologies with scrambling and faster decay have reduced its effectiveness.
Hardware-based acquisition utilizes interfaces such as FireWire, Thunderbolt, or PCIe to extract memory via DMA. Tools like Inception~\cite{inception2020} can read memory from live systems without OS mediation. However, DMA protections (e.g., Intel VT-d, Kernel DMA Protection) have largely mitigated these methods~\cite{microsoft_dma}.

\noindent In both cases, malware may evade detection by remapping physical addresses, populating decoy regions, or persisting in firmware to obscure traces during power cycles.

\noindent\textbf{Virtualization-Based Acquisition.}
Virtualized environments such as VMware, Hyper-V, and VirtualBox can suspend guest systems and serialize memory to disk, producing clean snapshots ideal for forensic analysis. However, the snapshot process may inadvertently notify guest applications through services like VMware Tools, giving malware an opportunity to erase traces before capture.

\noindent\textbf{Software-Based Acquisition.}
Software-driven acquisition relies on OS utilities and dedicated dumping tools. On Windows, memory can be captured using crash dump tools such as NotMyFault or LiveKD~\cite{russinovich_tools}. However, kernel-level rootkits may sanitize memory regions prior to dumping, and hardware-reserved areas can be omitted~\cite{dolan2007rootkits}. Additionally, this technique requires a forced crash and depends on pre-configured settings.

Alternative approaches include examining the Windows hibernation file (hiberfil.sys), though this process notifies running processes beforehand, allowing malware to erase data before capture~\cite{sans_hibernation}. Custom memory dumping tools operating within the OS face similar limitations: on a compromised system, attackers may tamper with the dumper or sanitize memory in real time.

\subsubsection{Trusted Acquisition Limitations}
\label{sec:backgroung:limitations}
In all memory acquisition categories, the attacker’s control over the execution context remains a barrier to fully trusted memory acquisition. If the OS, hypervisor, or firmware is compromised, no software- or virtualization-based acquisition method is inherently trustworthy \cite{rutkowska2006subverting}. Malware can hook memory APIs, supply forged data, or trigger anti-forensics routines. Accordingly, there is no purely software-based method that can guarantee secure and reliable memory acquisition.
Hardware-based mitigation techniques like DMA mitigations, memory encryption (e.g., Intel SGX), and TPM-backed attestation raise the bar but still do not guarantee forensic reliability in a fully adversarial setting, which may include supply chain compromise, malicious hardware modification, and hardware based implants \cite{sgx_whitepaper,parno_attestation}. Full forensic reliability therefore requires early capture as well as hardware-rooted trust, which is a subject beyond the scope of this work. 

Despite these limitations, practically, memory forensics remains one of the most popular and effective approaches available to incident responders and forensic analysts \cite{case2017volatility}. It enables deep inspection of the system state, including inspection of volatile artifacts that are not recorded elsewhere, such as in-memory-only malware, encryption keys, and active process structures. Even under adversarial conditions, memory analysis often provides evidence (or an unexpected lack thereof) that supports broader investigations when correlated with disk, network, and log data \cite{case2017volatility}. 

Our framework acknowledges this limitation. While it improves visibility into the UEFI memory earlier in the boot process, it does not attempt to solve the fully-trusted acquisition problem, which requires integration with secure hardware, cryptographic attestation, or vendor-controlled firmware, all of which may still be compromised in a fully adversarial setting.

\subsection{The UEFI}
The UEFI specification, developed by the UEFI Forum~\cite{UEFIAbout}, defines platform firmware architecture and functionality. 
Replacing the legacy BIOS, which could not meet modern hardware and software demands, UEFI implements a modular and extensible design. 
This architecture provides a standardized pre-boot environment for system initialization and hardware configuration, offering enhanced capabilities including support for larger disk drives, faster boot times, hardware abstraction, and security mechanisms.
UEFI firmware is often called a `mini OS' because it initializes and interacts with hardware, opens network connections, writes to disk, loads drivers, exposes a shell for user interaction, and runs UEFI applications--all to prepare the computer for the main OS.
UEFI firmware is typically stored as an image on one or more SPI flash chips on the motherboard. 
This non-volatile storage ensures persistence across power cycles and supports updates through UEFI Capsule Updates~\cite{CapsuleUpdate}. While essential for reliability, this persistence also poses a security risk: once compromised, malicious implants can survive OS reinstalls, disk formatting, and even disk replacements, allowing attackers to maintain persistence.
The UEFI boot process is organized into multiple phases, each fulfilling a specific role in system initialization. The process begins with the Security (SEC) phase, which verifies firmware integrity and initializes a temporary execution environment. Next, the Pre-EFI Initialization (PEI) phase identifies and configures the main system memory. The Driver Execution Environment (DXE) phase, central to UEFI functionality, loads drivers and protocols to configure hardware and prepare the platform for the OS handoff. Due to the importance of this phase, we elaborate on it further in~\autoref{sec:background:dxe}. 
Following the DXE phase, the Boot Device Selection (BDS) phase identifies and loads the OS bootloader. Before the OS takes over, the Transient System Load (TSL) phase may occur, where a bootloader or an application like the UEFI shell prepares the system for the final control handoff to the OS. Finally, the Runtime (RT) phase maintains the UEFI runtime services that persist in the OS environment after boot, facilitating ongoing firmware-OS interaction.

\subsection{Driver Execution Environment (DXE)}
\label{sec:background:dxe}
The DXE phase is the linchpin of the UEFI boot process. 
It connects the early initialization performed in the PEI phase with the BDS phase, which is responsible for loading and launching the OS. 
During this phase, the computer system transitions from a minimal setup to a fully functional execution environment capable of supporting platform services.

\subsubsection{UEFI Services}
\begin{figure}[ht]
\includegraphics[width=\columnwidth]{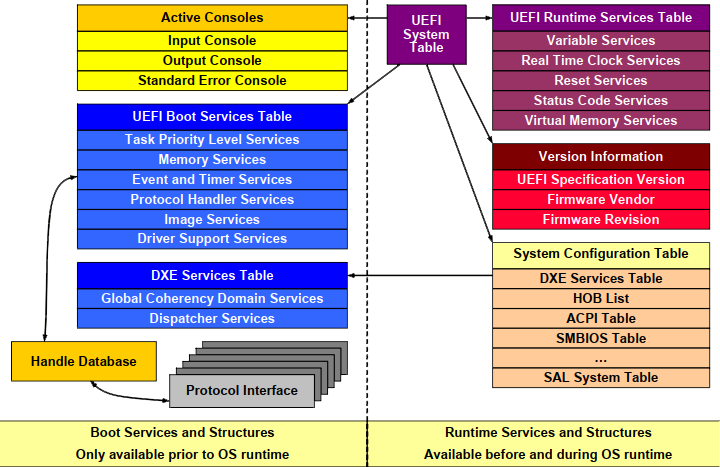}
\captionsetup{font=small, width=.9\columnwidth}
\caption{UEFI system table and related components \cite{uefi_pi_2023}.}
\label{fig:uefitables}
\end{figure}

UEFI services are divided into boot services, available only during platform initialization, and runtime services, which persist after OS handoff. 
The API for these services is organized into Service Tables (EFI System Table, EFI Boot Services Table, EFI Runtime Services Table, and EFI DXE Services Table), as depicted in~\autoref{fig:uefitables}.

\vspace{1.5ex}
\noindent\textbf{EFI System Table.}
The EFI System Table serves as the interface for accessing firmware services and interacting with higher-layer software. It is partially initialized during the PEI phase, where its basic structure is set up, and is fully populated during the DXE phase by the DXE Core. This table includes:

\begin{itemize}[topsep=0pt,noitemsep,leftmargin=*]
    
    \item Pointers to the EFI Boot Services Table and the EFI Runtime Services Table;
    
    \item A list of configuration tables identified by Globally Unique Identifiers (GUIDs), providing platform-specific or extended functionality; and 
    
    \item Metadata, such as the firmware vendor, firmware version, and UEFI specification revision.
    
\end{itemize}

\vspace{1.5ex}
\noindent\textbf{EFI Boot Services Table.}
The EFI Boot Services Table provides access to functions necessary for boot operations, including hardware initialization and OS handoff. In a shorthand convention, this table is known as the \texttt{gBS}. Key services include memory management, protocol management, event services, and timer services. Additional details about the \texttt{gBS} services are provided in Appendix~\ref{app:gBS}. 

\vspace{1.5ex}
\noindent\textbf{EFI Runtime Services Table.}
The services provided by this table remain available after the boot process is complete, exposing low-level services for the OS to use during its runtime. In a shorthand convention, this table is known as the \texttt{gRT}. Key runtime services include variable services, time services, and system reset services. Additional details about the \texttt{gRT} services are provided in Appendix~\ref{app:gRT}. 

\vspace{1.5ex}
\noindent\textbf{EFI DXE Services Table.}
The EFI DXE Services Table provides services specifically designed for the DXE phase of the UEFI boot process. These services enable the management of memory and I/O spaces, as well as the dispatching and coordination of DXE drivers. In a shorthand convention, this table is known as the \texttt{gDS}. Key services include memory space management, I/O management, driver dispatch, and firmware volume processing services. Additional details about the \texttt{gDS} services are provided in Appendix~\ref{app:gDS}.

\subsubsection{Image Loading}
UEFI images are categorized into Boot Service Drivers, which run during the boot phase; Runtime Drivers, which persist beyond boot; and Applications, which run when invoked. DXE drivers are typically embedded within the firmware image, but they can also be dynamically loaded from peripheral devices as Optional ROMs (OPROMs). Additional images, such as OS boot loaders and diagnostic utilities, can be loaded from the EFI System Partition (ESP) using the \texttt{gBS->LoadImage()} service and executed via \texttt{gBS->StartImage()}.

All UEFI images are associated with a GUID, a 128-bit unique identifier used to track and manage system resources. For OPROMs, the GUID is derived from the device's Vendor ID and Device ID. For applications, the GUID may be associated with the file path from which the image was loaded. Regardless of source or type, all loaded images conform to the Portable Executable/Common Object File Format (PE/COFF), sharing the same in-memory structure.

\subsubsection{ExitBootServices}
\label{sec:backgroung:ExitBootServices}
The DXE phase concludes with the \textit{ExitBootServices} event, which triggers the transition to the OS. This marks the completion of hardware and firmware initialization, ensuring that the platform is fully prepared for OS operation while reducing the firmware's role to retaining only runtime services for use during OS runtime.
At this point, the \texttt{ExitBootServices()} function is called to terminate boot services: the EFI Boot Services Table and the EFI DXE Services Table are terminated, resources are released, and the firmware is prevented from interfering with the OS’s management of hardware and memory.
Conversely, the EFI Runtime Services Table persists.
While the Boot and DXE Services Tables are terminated and the Runtime Services Table is preserved, the EFI System Table adopts an interim approach: after \texttt{ExitBootServices()} is called, only select entries such as pointers to the Runtime Services Table and configuration tables remain valid, enabling continued OS-firmware interaction.

\subsection{UEFI Bootkits and Attack Techniques}

\subsubsection{Malicious Image Loading in the UEFI}
The modularity of UEFI image loading introduces potential attack vectors through unauthorized image execution during boot. While Secure Boot~\cite{eclypsium_secure_boot_dbx} verifies digital signatures of DXE drivers, sophisticated bootkits have demonstrated the ability to bypass these protections. \autoref{tab:ImageLoading} outlines locations from which attackers can load malicious code through UEFI's image loading infrastructure.

\begin{table}[!ht]
\caption{Malicious image loading by UEFI bootkits from the ESP, SPI flash, OPROM, or UEFI shell.}
\centering
\footnotesize
\setlength{\tabcolsep}{3pt}
\renewcommand{\arraystretch}{1.05}

\resizebox{\linewidth}{!}{
\begin{tabular}{l||l|c|c|c|c|l}
\hline
\textbf{Bootkit} & \textbf{Type} & \textbf{ESP} & \textbf{SPI} & \textbf{OP.} & \textbf{Shell} & \textbf{Ref.}\\ 
\hline \hline
DarkSeaSkies     & Attack    &  $\scalebox{1.5}{$\circ$}$ & $\scalebox{1.5}{$\bullet$}$ & $\scalebox{1.5}{$\circ$}$ & $\scalebox{1.5}{$\circ$}$ & \cite{firmwaresecurity_darkmatter}\\
DerStrake     & Attack    & $\scalebox{1.5}{$\circ$}$ & $\scalebox{1.5}{$\bullet$}$ & $\scalebox{1.5}{$\circ$}$ & $\scalebox{1.5}{$\circ$}$ & \cite{wikileaks_derstarke_v1_4}\\
DreamBoot     & PoC    & $\scalebox{1.5}{$\bullet$}$ & $\scalebox{1.5}{$\circ$}$ & $\scalebox{1.5}{$\circ$}$ & $\scalebox{1.5}{$\circ$}$ & \cite{dreamboot2013}\\
Thunderstrike & Attack & $\scalebox{1.5}{$\circ$}$ & $\scalebox{1.5}{$\circ$}$ & $\scalebox{1.5}{$\bullet$}$ & $\scalebox{1.5}{$\circ$}$ & \cite{Thunderstrike}\\
Thunderstrike2 & PoC   & $\scalebox{1.5}{$\circ$}$ & $\scalebox{1.5}{$\bullet$}$ & $\scalebox{1.5}{$\bullet$}$ & $\scalebox{1.5}{$\circ$}$ & \cite{thunderstrike2}\\
VectorEDK      & Attack& $\scalebox{1.5}{$\circ$}$ & $\scalebox{1.5}{$\bullet$}$ & $\scalebox{1.5}{$\circ$}$ & $\scalebox{1.5}{$\circ$}$ & \cite{vector-edk}\\
LightEater & PoC   & $\scalebox{1.5}{$\circ$}$ & $\scalebox{1.5}{$\bullet$}$ & $\scalebox{1.5}{$\circ$}$ & $\scalebox{1.5}{$\circ$}$ & \cite{LightEater}\\
LoJax          & Attack & $\scalebox{1.5}{$\circ$}$ & $\scalebox{1.5}{$\bullet$}$ & $\scalebox{1.5}{$\circ$}$ & $\scalebox{1.5}{$\circ$}$ & \cite{LoJax}\\
MosaicRegressor& Attack & $\scalebox{1.5}{$\circ$}$ & $\scalebox{1.5}{$\bullet$}$ & $\scalebox{1.5}{$\circ$}$ & $\scalebox{1.5}{$\circ$}$ & \cite{MosaicRegressor}\\
umap          & PoC   & $\scalebox{1.5}{$\bullet$}$ & $\scalebox{1.5}{$\circ$}$ & $\scalebox{1.5}{$\circ$}$ & $\scalebox{1.5}{$\circ$}$ & \cite{umap2020} \\
ESPecter       & Attack & $\scalebox{1.5}{$\bullet$}$ & $\scalebox{1.5}{$\circ$}$ & $\scalebox{1.5}{$\circ$}$ & $\scalebox{1.5}{$\circ$}$ & \cite{especter}\\
FinSpy         & Attack & $\scalebox{1.5}{$\bullet$}$ & $\scalebox{1.5}{$\circ$}$ & $\scalebox{1.5}{$\circ$}$ & $\scalebox{1.5}{$\circ$}$ & \cite{FinSpy}\\
MoonBounce     & Attack & $\scalebox{1.5}{$\circ$}$ & $\scalebox{1.5}{$\bullet$}$ & $\scalebox{1.5}{$\circ$}$ & $\scalebox{1.5}{$\circ$}$ & \cite{MoonBounce_2022}\\
CosmicStrand   & Attack & $\scalebox{1.5}{$\circ$}$ & $\scalebox{1.5}{$\bullet$}$ & $\scalebox{1.5}{$\circ$}$ & $\scalebox{1.5}{$\circ$}$ & \cite{CosmicStrand_2022}\\
BlackLotus     & Attack & $\scalebox{1.5}{$\bullet$}$ & $\scalebox{1.5}{$\circ$}$ & $\scalebox{1.5}{$\circ$}$ & $\scalebox{1.5}{$\circ$}$ & \cite{blacklotus}\\
NotPetyaAgain  & PoC   & $\scalebox{1.5}{$\circ$}$ & $\scalebox{1.5}{$\circ$}$ & $\scalebox{1.5}{$\circ$}$ & $\scalebox{1.5}{$\bullet$}$ & \cite{NotPetyaAgain}\\
EfiGuard       & PoC   & $\scalebox{1.5}{$\bullet$}$ & $\scalebox{1.5}{$\circ$}$ & $\scalebox{1.5}{$\circ$}$ & $\scalebox{1.5}{$\circ$}$ & \cite{EfiGuard}\\
Glupteba       & Attack & $\scalebox{1.5}{$\bullet$}$ & $\scalebox{1.5}{$\circ$}$ & $\scalebox{1.5}{$\circ$}$ & $\scalebox{1.5}{$\circ$}$ & \cite{Glupteba_2024}\\
Bootkitty      & PoC    & $\scalebox{1.5}{$\bullet$}$ & $\scalebox{1.5}{$\circ$}$ & $\scalebox{1.5}{$\circ$}$ & $\scalebox{1.5}{$\circ$}$ & \cite{bootkitty}\\
\hline\hline
\multicolumn{7}{l}{\footnotesize Legend: $\scalebox{1.5}{$\bullet$}$ image loaded or $\scalebox{1.5}{$\circ$}$ not loaded from that source.} \\
\multicolumn{7}{l}{\footnotesize \textbf{PoC}: described online but not seen in the wild.} \\
\multicolumn{7}{l}{\footnotesize \textbf{Attack}: seen implemented in the wild.}\\
\hline \hline
\end{tabular}
}
\label{tab:ImageLoading}
\end{table}

\subsubsection{Hooking in the UEFI}
Hooking is a code manipulation technique that hijacks the program execution flow and redirects it to enable the execution of an attacker's code. 
The typical approach involves modifying function pointers or instructions using two primary techniques: function pointer hooking and inline hooking. 
A summary of bootkits that employ UEFI-level hooks is provided in Appendix~\ref{app:bootkitsHookingSummary}.

\begin{figure}[ht]
\includegraphics[width=\columnwidth]{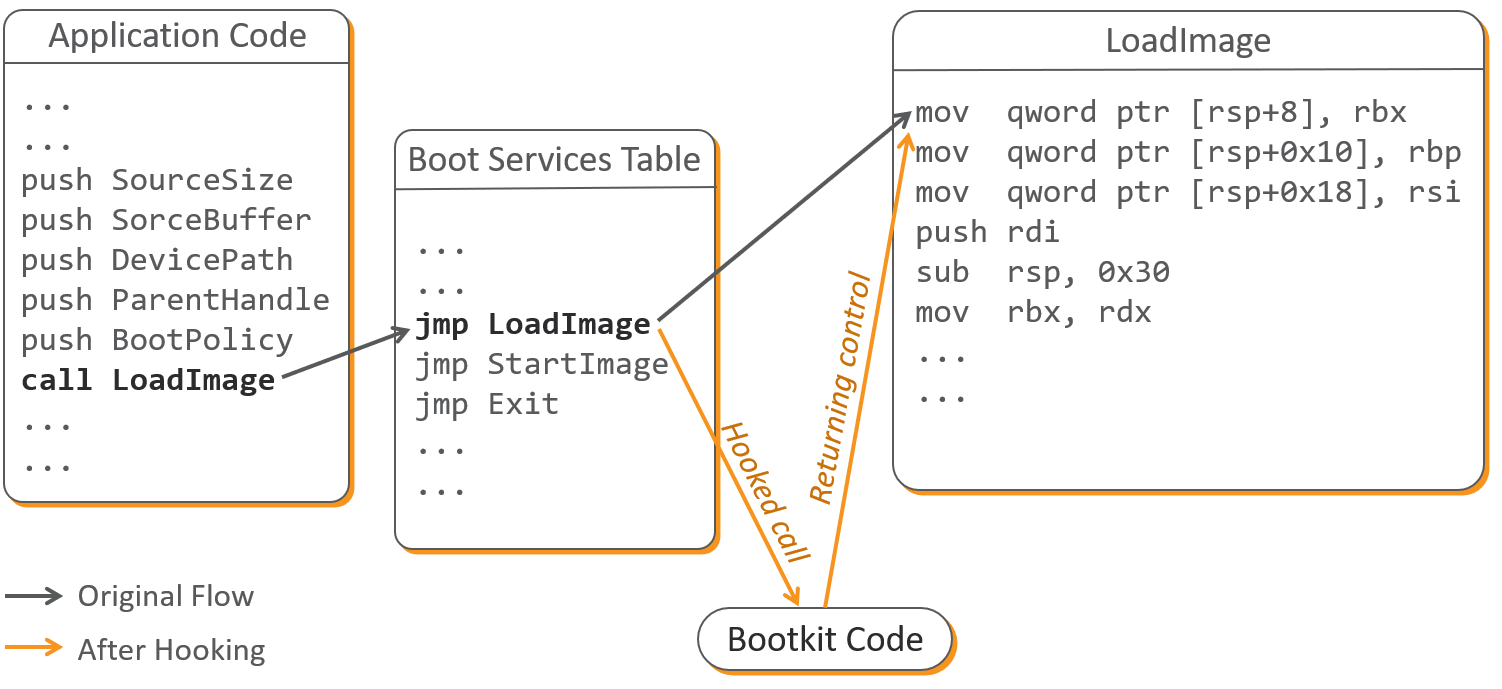}
\captionsetup{font=small, width=0.9\columnwidth}
\caption{Function pointer hooking.}
\label{fig:tradHooking}
\end{figure}

\vspace{1.5ex}
\noindent\textbf{Function Pointer Hooking. }
Function pointer hooking replaces a legitimate function address with the address of attacker-controlled code. 
When the system calls the hooked function, execution is redirected to the malicious code, allowing attackers to intercept calls and execute their own code before or instead of the original functionality (see \autoref{fig:tradHooking}).
For example, Glupteba's UEFI bootkit~\cite{Glupteba_2024} replaces the \texttt{LoadImage} entry in the Boot Services Table with a malicious handler, causing every image load during boot to pass through the bootkit. This allows it to patch subsequent components in the boot chain, including the Windows Boot Manager, ultimately disabling security mechanisms during startup.

\begin{figure}[ht]
\includegraphics[width=\columnwidth]{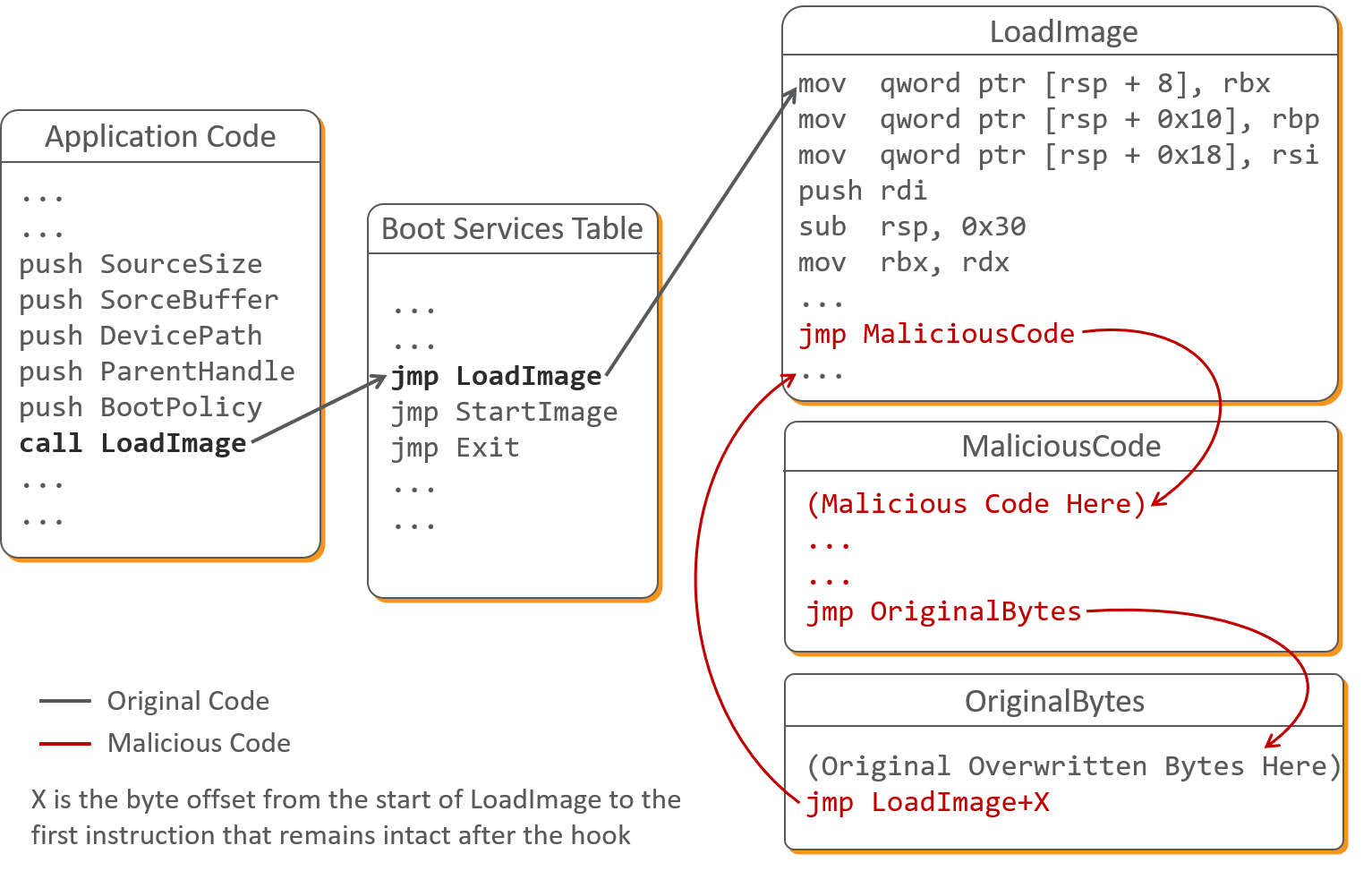}
\captionsetup{font=small, width=0.9\columnwidth}
\caption{Inline hooking.}
\label{fig:InlineHooking}
\end{figure}

\noindent\textbf{Inline Hooking. }
Inline hooking directly alters machine code at function entry points by replacing the first instructions with a \texttt{call} or \texttt{jump} instruction that redirects execution to attacker-controlled code. The attacker's code can execute before, after, or instead of the original function, typically using a trampoline mechanism to preserve the overwritten instructions and allow normal execution to continue (see~\autoref{fig:InlineHooking}).

In the UEFI context, the MoonBounce bootkit~\cite{MoonBounce_2022} implements inline hooking by patching the entry points of \texttt{AllocatePool}, \texttt{CreateEventEx}, and \texttt{ExitBootServices}. Hooking \texttt{AllocatePool} allows interception of memory allocation requests during DXE; \texttt{CreateEventEx} enables triggering malicious actions at specific boot events; and \texttt{ExitBootServices}, invoked just before OS handoff, allows the attacker to inject code into the loader and subvert the kernel before boot completes.

\subsubsection{Control-Flow Hijacking via Gadget Chains}
While existing UEFI bootkits primarily rely on persistent modifications such as service table hooks or malicious image loading, an adversary with code execution capabilities during DXE could employ gadget-based techniques similar to Return-Oriented Programming (ROP). In this approach, the attacker reuses short instruction sequences (gadgets) from existing PE/COFF images, chaining them together using manipulated stack contents or non-standard control transfers.

%% file: Sections/ThreatModel.tex
\section{Threat Model}
\label{sec:threat_model}
\noindent\textbf{Attacker Capabilities.}
We consider an adversary who can execute code in the UEFI environment during DXE or later pre-OS stages.
The attacker operates within the UEFI model and may call Boot Services and Runtime Services.
The attacker can load components, modify volatile firmware state, change control flow, or influence the boot process.
We define these capabilities by the actions the attacker can perform rather than by specific techniques, since firmware threats evolve and often combine several methods.
The threat model captures the behavior seen in existing UEFI bootkits, but is not limited to specific families.
The attacker may also prepare structured data in writable firmware memory, including values that reference code locations indirectly, such as offsets relative to loaded images. These values can be resolved at runtime once image bases are known. Our framework considers such preparation behaviors as part of general attacker capabilities, without assuming any specific exploit technique.

\vspace{1.5ex}
\noindent\textbf{System Setting.}
The platform follows the standard UEFI boot flow.
During DXE, the firmware loads drivers, initializes hardware, and builds the System Table and service tables.
PE/COFF images are placed in memory at this stage.
Our framework analyzes a snapshot taken during this phase and inspects these structures offline.

\vspace{1.5ex}
\noindent\textbf{Defender Goals.}
The defender uses our framework to detect violations of basic structural and ownership properties in UEFI memory.
Examples include service table entries that do not point to valid code in legitimate drivers, malformed or unexpected images, or writable regions that show abnormal control-flow patterns.
The goal is to expose modifications that occur before OS initialization and may not be visible later.
The framework supports forensic inspection of a captured snapshot and does not provide runtime protection or attestation capabilities.

\noindent\textbf{Trust Assumptions.}
We assume that the memory acquisition component runs during the UEFI phase and records a consistent snapshot.
It is trusted to record memory accurately, although it shares the same privilege level as the attacker.
We do not claim that the snapshot is immune to tampering by a fully capable attacker.
As in OS memory forensics, some evidence may be removed before capture.
Our analysis focuses on inconsistencies that remain detectable.
Memory must be acquired from within the UEFI environment, since earlier or higher-privilege acquisition methods are not feasible in practice.
We also assume that core UEFI structures must remain present and well-formed for the system to boot.
The attacker may modify entries, but cannot remove these structures entirely.
Our detection methods rely on this constraint by checking the consistency of pointers, images, and memory layout.

%% file: Sections/Methodology.tex
\section{Our Framework} 
\label{sec:methodology}
In this section, we describe our framework for UEFI memory analysis, which consists of two main components: \textit{UEFIMemDump}, responsible for memory acquisition, and \textit{UEFIDumpAnalysis}, which is comprised of several analysis modules that process the dump. These components enable systematic examination of UEFI memory during the boot phase, facilitating the detection of threats before OS initialization. 

\begin{figure*}[!htb]
\centering
\includegraphics[scale=0.52]{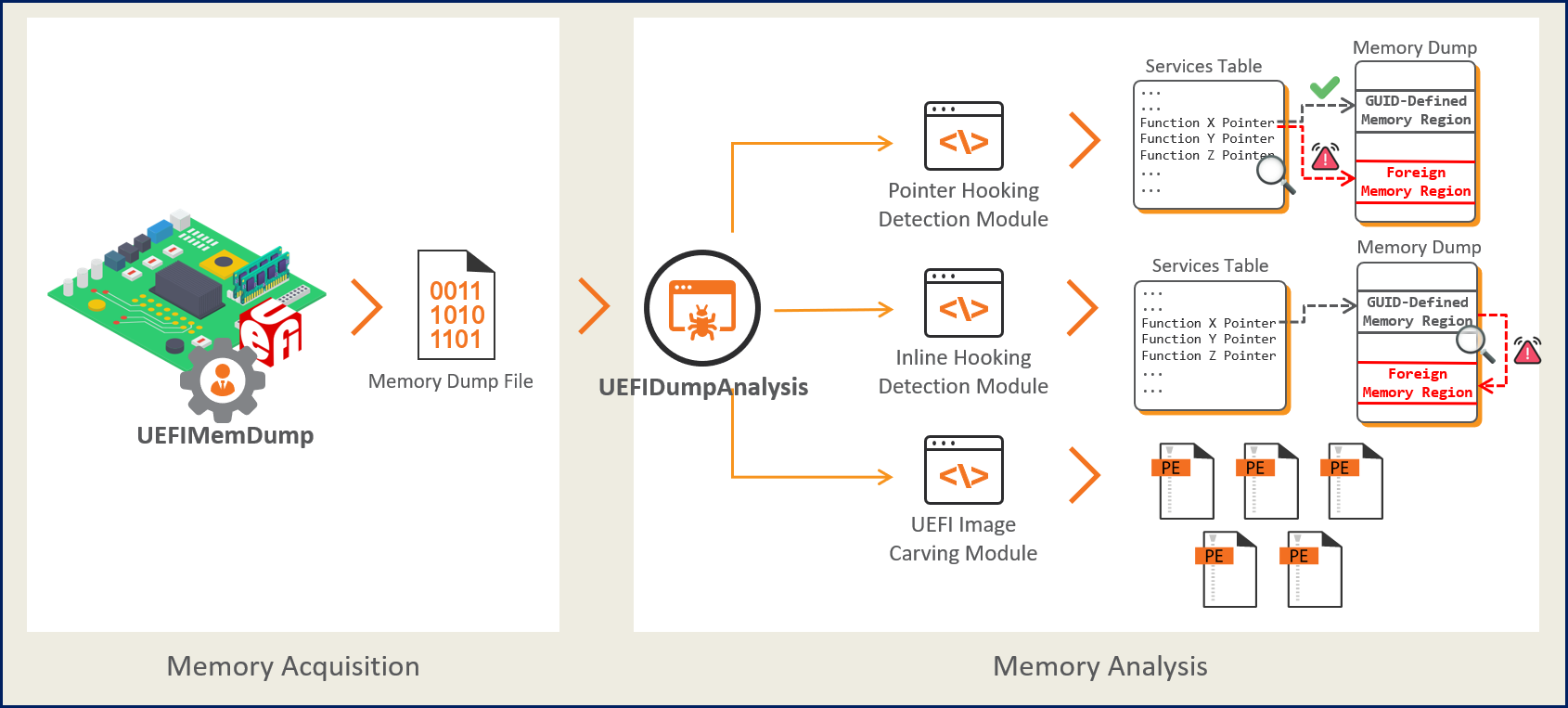}
\caption{Framework architecture.}
\label{fig:framework}

\end{figure*}

As can be seen in \autoref{fig:framework}, the framework's components are implemented in a two-stage approach: memory collection is performed via the \textit{UEFIMemDump} component and subsequent analysis is conducted through one or more modules of the \textit{UEFIDumpAnalysis} component. Together they serve as the basis of our framework's forensic capabilities, enabling it to identify threats in the UEFI environment.
Since our analysis relies on standard UEFI runtime and boot-time data structures as defined in the UEFI specification, our framework is inherently generalizable across firmware implementations that conform to the standard. As a result, our framework is vendor-agnostic and can operate reliably across a broad range of UEFI-compliant platforms. The analysis relies only on the memory structures present in the memory snapshot and does not require symbols or source level metadata.

\subsection{Memory Acquisition with UEFIMemDump}
The framework's core memory acquisition capability derives from \textit{UEFIMemDump}, which enables memory acquisition during the boot process. Its primary function is to collect complete memory snapshots of the UEFI environment, providing visibility into both persistent and transient memory regions. Implemented within the EDK II ecosystem~\cite{edk2} as both a DXE driver and a UEFI shell application, it ensures cross-platform compatibility while capturing critical memory regions that are typically inaccessible in post-boot OS environments.
We chose to implement this memory dumping utility in both forms for practical reasons: while a DXE driver may be suitable for virtual environments, it may not be practical to compile it into the UEFI firmware of a physical machine, where a UEFI shell application is more suitable and enables forensic investigation without requiring firmware modification.
The memory acquisition process is performed as follows:
\begin{itemize} [itemsep=0.1em,topsep=2pt,leftmargin=*]
    \item \textbf{Memory Mapping}: Upon loading, \textit{UEFIMemDump} invokes \texttt{GetMemoryMap()} to construct a detailed map of system memory regions, ensuring complete coverage for analysis; 
    \item \textbf{Memory Acquisition}: Next, the memory is copied bit by bit. The collection process targets the mapped memory regions; and 
    \item \textbf{Data Storage}: The acquired memory dump is written to a file in a raw binary format, which is analyzed in the subsequent memory analysis stage.
 \end{itemize}

\subsection{Dump Analysis with UEFIDumpAnalysis}
\textit{UEFIDumpAnalysis} serves as the analytical component of our forensic framework, enabling the examination of UEFI memory captured by \textit{UEFIMemDump}. As shown in \autoref{fig:framework}, \textit{UEFIDumpAnalysis} is implemented as an extendable collection of modules, each focusing on the detection of UEFI-specific threats commonly employed by modern bootkits. In our PoC implementation, we developed two modules focused on hooking detection, one module for extracting images from memory, and an additional module for detecting chained control-flow patterns based on gadget references.

\subsubsection{Parsing UEFI Data Structures}
The analysis starts with the extraction and interpretation of the UEFI data structures found in the memory, such as the EFI Boot Services Table, the EFI Runtime Services Table, the EFI DXE Services Table, and the loaded images. As mentioned previously, these structures encapsulate both metadata and function pointers (where relevant) that govern system operations across the boot and runtime phases. 
As part of our research, we were able to identify and interpret the in-memory structures.

\begin{figure}[ht]
\centering
\fbox{
\includegraphics[trim=1.5cm 0cm 1.5cm 0cm, clip,scale=0.18]{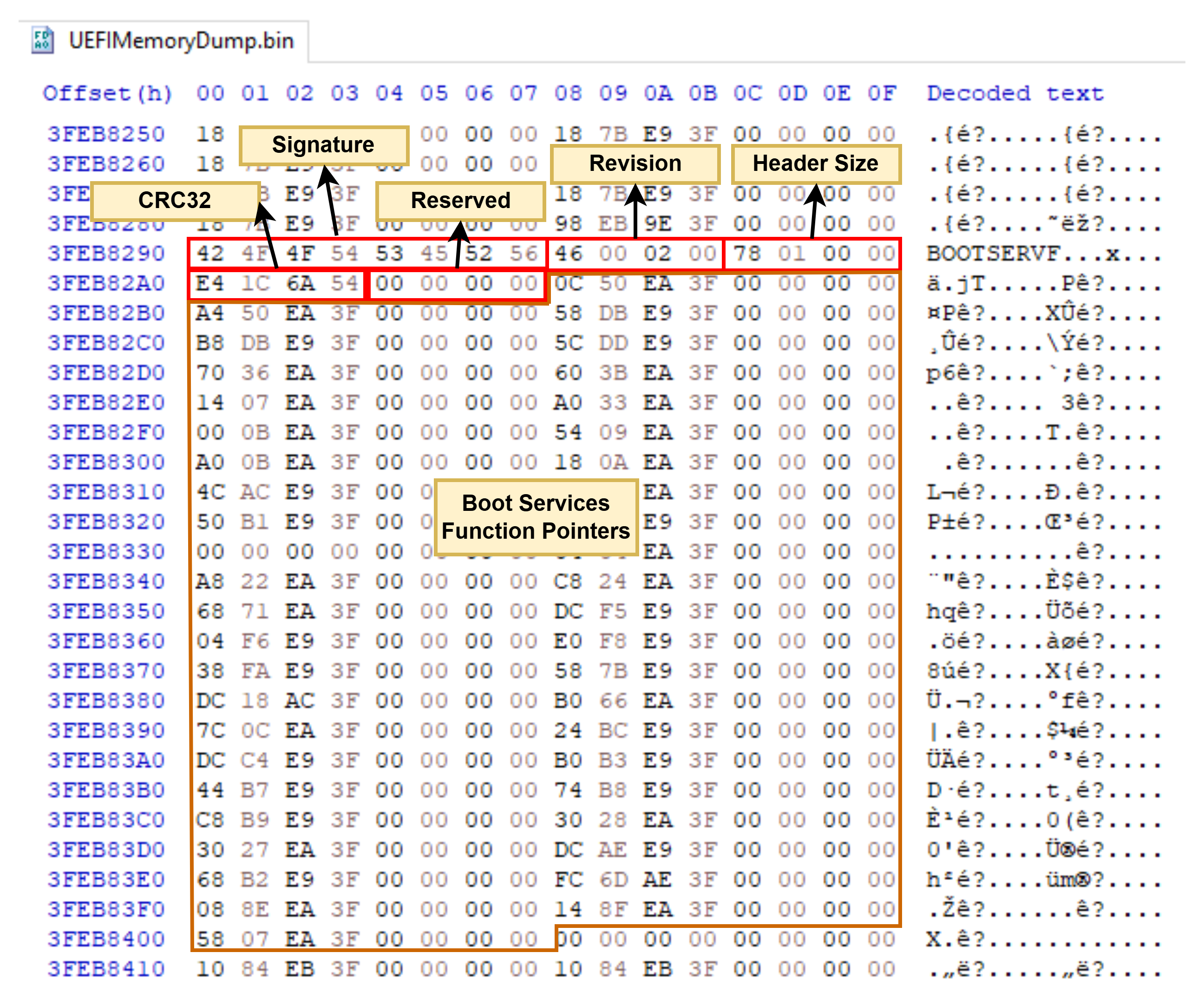}
}
\caption{The in-memory fields of the EFI Boot Services Table, containing metadata and function pointer addresses for the services.}
\label{fig:BootServ}
\end{figure}

\begin{figure}[ht]
\centering
\fbox{
\includegraphics[trim=1.2cm 0cm 1.7cm 0cm, clip, scale=0.27]{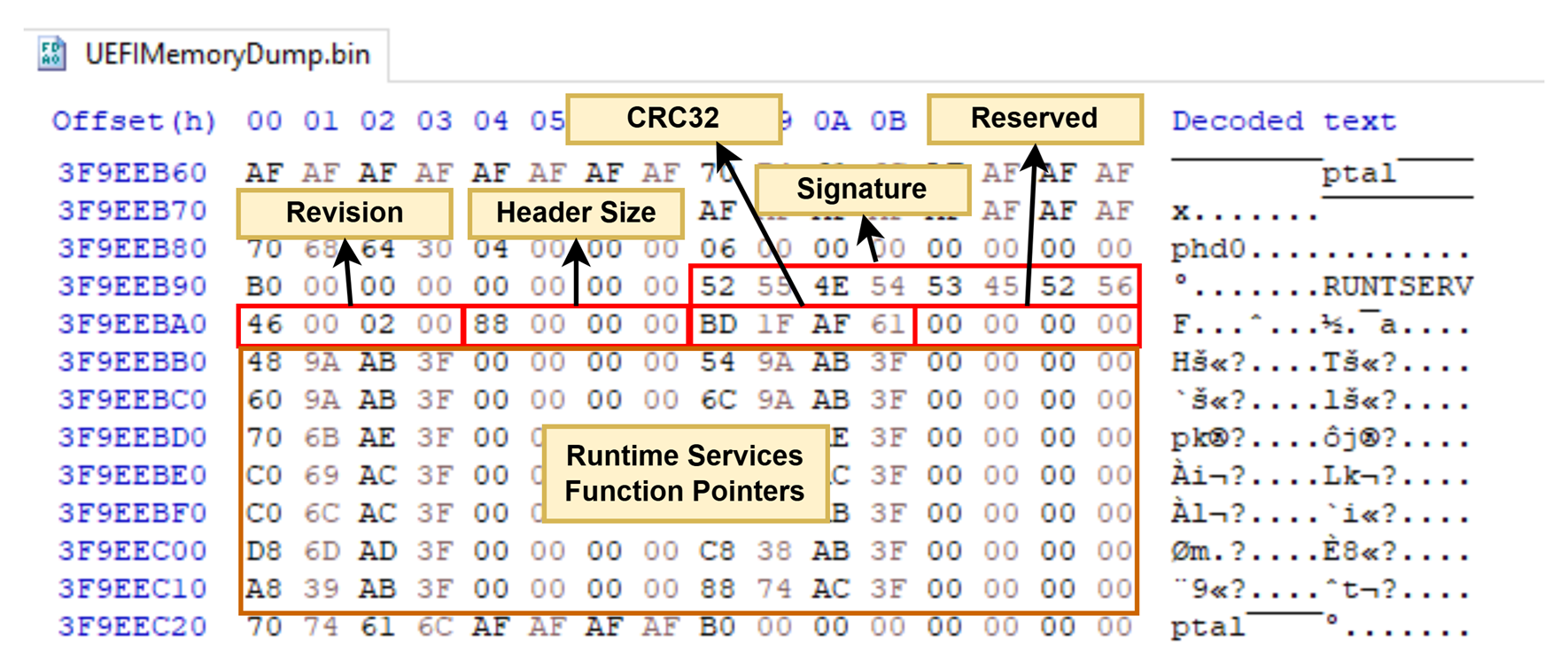}
}
\caption{The in-memory fields of the EFI Runtime Services Table, containing metadata and function pointer addresses for the services.}
\label{fig:RunServ}
\end{figure}

As seen in \autoref{fig:BootServ}, the \textit{EFI Boot Services Table} structure resides in memory starting with the signature \texttt{b'BOOTSERV'}. 
It contains metadata fields such as CRC32 checksum, revision, and header size, followed by an array of function pointers pointing to boot services such as \texttt{CreateEvent}, \texttt{AllocatePages}, and \texttt{ExitBootServices}.
Similarly, the beginning of the \textit{EFI Runtime Services Table} structure in memory is marked by the signature \texttt{b'RUNTSERV'}, as seen in \autoref{fig:RunServ}. Like the EFI Boot Services Table structure, this structure includes metadata fields ensuring structural integrity and compatibility validation. 

In addition to the metadata, the EFI Runtime Services Table contains function pointers to runtime services such as \texttt{GetVariable}, \texttt{SetVariable}, and \texttt{ResetSystem}, which remain accessible after the OS has been loaded.
The \textit{EFI DXE Services Table}, which is presented in Appendix~\ref{app:structureparsing}, in Figure~\ref{fig:DxeServ}, is another critical structure located and analyzed by the \textit{UEFIDumpAnalysis} component. Starting with the signature \texttt{b'DXE\_SERV'}, it provides essential services for the DXE phase, including memory space management, I/O space management, and firmware volume processing. The table contains metadata fields similar to other UEFI tables, ensuring consistency and validity, as well as function pointers to DXE-phase-specific services such as \texttt{AddMemorySpace}, \texttt{AllocateIoSpace}, and \texttt{Dispatch}.

In addition to these service tables, loaded UEFI images--including runtime drivers, boot service drivers, and UEFI applications--are identified with the help of the \texttt{b'ldri'} signature, marking image headers, as seen in Appendix~\ref{app:structureparsing} in Figure~\ref{fig:DriverImage}. This signature enables the extraction of relevant metadata, including the image's base address, size, and either a file path or GUID reference. 
The image base address indicates where the loader placed the UEFI image file in memory during loading.
The extraction and parsing of these data structures form the common foundation for the detection modules described below.

\subsubsection{UEFI Image Carving Module}
The \textit{UEFI Image Carving Module} focuses on extracting PE/COFF files from memory dumps. This module extracts the images loaded to memory by parsing the image base address and image size fields residing in the \texttt{b'ldri'} structures.
Unlike traditional methods that rely on scanning for \texttt{'MZ'} headers, this module uses the structured information within the \texttt{b'ldri'} entries to identify valid PE files loaded by the UEFI \texttt{LoadImage} boot service. Extracted images, including drivers and other executables, are saved to a designated output folder, with filenames derived from their GUIDs or file paths. These carved  files can then be subjected to additional analysis, such as applying YARA rules to detect known malicious patterns or identify anomalous characteristics. 

\subsubsection{Gadget-Based Control-Flow Detection Module}
\label{sec:rop_detection}

This module identifies sequences of values in firmware memory that reference code locations inside DXE images. All carved PE/COFF drivers are processed with Ropper~\cite{ropper} to extract short instruction sequences (gadgets) that terminate in a \texttt{ret} or indirect branch. For each image we record gadget offsets and the corresponding runtime addresses (image\_base + offset), and build two lookup tables: an offset map and a runtime map.

When scanning memory the module reads 64-bit words and evaluates them in two modes. In the resolved-address mode each value is treated as an absolute pointer and matched against the runtime map. In the offset mode values are interpreted as gadget offsets and matched against the offset map. Both modes reflect realistic DXE-level adversaries who can prepare structured data in writable memory either as unresolved offsets or as absolute addresses once image bases are known.

Simple contiguous-run detection is applied to identify regions with a high density of gadget references. Runs that exceed minimum length and show consistent provider attribution are flagged as suspicious and reported with both their original and resolved sequences.

\subsubsection{Function Pointer Hooking Detection Module}
The \textit{Function Pointer Hooking Detection Module} initiates the extraction and analysis of function pointers from relevant service tables such as the Boot Services Table, Runtime Services Table, and DXE Services Table, since they can be victims of hooking.
The entries in each UEFI-compliant service table include pointers to service functions. 
More specifically, each entry in the service table is a pointer to the DXE driver where the service is implemented.

To associate the service name with the respective pointer, the service tables follow a strict sequential order that is maintained across UEFI-compliant implementations, where each number in the table entry corresponds to a specific service function.
This order makes the extraction and analysis of function pointers straightforward: first, function names are mapped to their pointers in the corresponding service table.
Next, each service pointer undergoes validation to confirm that it resides within the expected memory range associated with its driver’s GUID.
If a service pointer in a service table references an address outside the memory region defined by the GUID of the driver responsible for implementing that service, it is flagged as suspicious, as it violates the expected mapping between the function and its originating driver. 
Upon detecting such anomaly, the \textit{Function Pointer Hooking Detection Module} records the relevant metadata, including the GUID, the driver's memory region, and, when available, the file path of the driver owning the memory region where the pointer was found; this enables analysts to trace the anomaly back to its source. 

\subsubsection{Inline Hooking Detection Module}
Since inline hooking is typically accomplished by overriding the code of the target function, the \textit{Inline Hooking Detection Module} begins its analysis by disassembling the code at the addresses referenced by service function pointers, using the Capstone disassembler \cite{capstone}.
The analysis specifically targets \texttt{jmp} and \texttt{call} instructions within the function prologue, as these are commonly exploited to hijack the execution flow in inline hooking. 
In this context, when the module detects a \texttt{call} or \texttt{jmp} instruction (including other variations of jump, such as \texttt{jnz} and \texttt{je}) redirecting the program flow to an unexpected memory range, it flags the instance as a potential inline hook.
It also records relevant metadata, including the suspected function name, hook address, target address, and the associated driver's GUID or file path. 

%% file: Sections/Evaluation.tex
\section{Evaluation}
\label{sec:evaluation}

As described in Section~\ref{sec:threat_model}, there are currently 12 documented UEFI bootkit attacks in the wild, along with numerous additional proof-of-concept implementations. These threats can be categorized based on their core techniques, with the most prevalent being function hooking (\autoref{tab:hooks}) and malicious image loading (\autoref{tab:ImageLoading}). In addition to these documented behaviors, firmware adversaries with DXE-level execution are also capable of preparing structured control-flow data in writable memory\cite{ropbin, ropbinsmm}, which motivates our inclusion of a control-flow analysis module. To ensure the framework provides generic detection capabilities that identify techniques rather than specific samples, we selected representative test cases covering these fundamental behaviors. 

We evaluated the framework using a controlled testbed environment where we executed attacks covering the two documented behaviors and added synthetic control-flow patterns that reflect actions a DXE-capable adversary is able to perform. Our test cases included implementations of techniques used by real-world threats such as Glupteba \cite{Glupteba_2024}, MoonBounce \cite{MoonBounce_2022}, and CosmicStrand \cite{CosmicStrand_2022}, along with publicly available proof-of-concept bootkits including EfiGuard \cite{EfiGuard} and Thunderstrike \cite{Thunderstrike}. This selection ensures coverage of the core documented behaviors while also demonstrating the framework’s ability to detect additional control-flow irregularities that may arise from adversaries operating within the UEFI environment.

\subsection{UEFIMemDump}
We implemented our memory dumping utility as both a DXE driver and a UEFI application. To perform an evaluation of the dumper that encompasses both the physical and virtual environments, we evaluated the DXE driver version of \textit{UEFIMemDump} in a virtual environment and the UEFI application version of \textit{UEFIMemDump} on a physical machine.
In both setups we were able to obtain non-corrupted UEFI memory dumps and save them as a raw binary file.

\subsubsection{Virtual Setup}
For our virtual environment, we utilized TianoCore EFI Development Kit II (EDK II), a popular open-source UEFI specification implementation maintained by the TianoCore community~\cite{edk2}. 
This implementation provides tools for firmware creation, testing, and validation while ensuring UEFI compliance. 
The experiments were conducted in a virtualized QEMU~\cite{qemu} environment, configured with 1 GB RAM, 1 CPU core, and an NTFS-formatted virtual disk (VHD) running Windows 11, and a FAT32-formatted 4 GB virtual USB storage device used as removable media. The EDK II image served as the system UEFI firmware image, with \textit{UEFIMemDump} embedded as a custom DXE driver, enabling memory acquisition during the UEFI boot process, immediately before the \textit{ExitBootServices} event is triggered.
In this setup, the DXE driver of the dumper was configured to write memory snapshots to the virtual USB storage device.

\subsubsection{Physical Setup}
We conducted experiments on two physical laptops with distinct platforms. 
The first system was a System76 Adder WS running Ubuntu 22.04 LTS with a 14th Gen Intel Core i9 processor and Coreboot firmware. 
The second system was a Lenovo ThinkPad T14 Gen4 running Windows 11 Enterprise with a 13th Gen Intel Core i7 processor and Lenovo UEFI firmware.

In this setup, the UEFI shell application version of the dumper was configured to write memory snapshots directly to an external USB SSD connected to each laptop. To assess scalability, we tested the framework on systems equipped with 32 GB of RAM and successfully acquired the full memory without any issues. When writing to the external USB SSD device, the acquisition process was completed in approximately two minutes. These results demonstrate that our framework reliably supports large memory configurations and is well-suited for practical use in real-world forensic investigations.

\subsection{UEFIDumpAnalysis}
After demonstrating the ability to obtain a memory snapshot, we proceeded with the evaluation of \textit{UEFIDumpAnalysis}. We executed the attacks in the virtual setup, where we could freely implement the attack scenarios as described in the subsections that follow. We introduced various attack scenarios using malicious DXE drivers, each assigned a unique GUID to prevent conflicts with existing system components. After executing each attack, we obtained memory snapshots with the DXE driver version of the \textit{UEFIMemDump} and ran the detection modules of the \texttt{UEFIDumpAnalysis} offline. 

\subsection{Gadget-based Control-Flow Detection Module}
\label{sec:rop_detection}
We evaluate both detection modes using PoC DXE drivers that construct synthetic sequences in writable memory. The provider image is \texttt{UsbBusDxe.efi}, and its gadgets are extracted with Ropper~\cite{ropper}. In the offset mode the PoC driver writes only gadget offsets to a buffer. The detector resolves these offsets by matching them against gadget tables from all carved drivers and identifies a run containing eight gadget entries. The result appears in Appendix~\ref{app:detection}, Figure~\ref{fig:gadgets_offets}. In the resolved-address mode the PoC driver discovers the runtime base of \texttt{UsbBusDxe.efi} and writes absolute gadget entry addresses. The detector recovers eleven gadget entries with full provider consistency. The result appears in Appendix~\ref{app:detection}, Figure~\ref{fig:gadgets_resolved}. These experiments show that the module detects both unresolved and execution-ready control-flow sequences using only the memory snapshot.

\subsection{Function Pointer Hooking Detection}
This module was applied on the memory dump to detect malicious modifications of function pointers in the Boot Services Table, Runtime Services Table, and DXE Services Table, identifying anomalies caused by pointer redirection in the scenarios described below.

\subsubsection{EfiGuard and Glupteba - Hooking via UEFI Driver and Loader in the ESP}
\mbox{}\\
\mbox{}\\
\textbf{Execution of the Attack:}  
EfiGuard \cite{EfiGuard} is an open-source UEFI bootkit designed to bypass Windows kernel protections such as PatchGuard and DSE. We obtained its source code and implemented an attack scenario involving its hooking routine. This attack leverages two key components: a UEFI driver (\texttt{EfiGuardDxe.efi}) and a loader (\texttt{Loader.efi}), both placed in the ESP.
During the boot process, the malicious \texttt{Loader.efi} replaces the legitimate Windows Boot Manager (\texttt{bootmgfw.efi}) to ensure early execution of the \texttt{EfiGuardDxe.efi} driver. The driver hooks critical functions in the Boot Services Table and Runtime Services Table. Specifically, the \texttt{LoadImage} function is hooked to intercept boot-time operations, enabling manipulation of kernel structures to disable PatchGuard, and the \texttt{SetVariable} function is also hooked to establish a runtime backdoor, facilitating kernel memory operations from the user space. To maintain integrity, the CRC32 checksum of the service tables is recalculated after the modifications.
The Glupteba\cite{Glupteba_2024} UEFI bootkit, which is derived from EfiGuard, modifies this behavior by hooking just the \texttt{LoadImage} function, leaving \texttt{SetVariable} unaltered. Therefore, the detection approach (which is able to detect the hooking of \texttt{LoadImage} in the style of EfiGuard) will also be capable of detecting Glupteba.

\begin{figure}[ht]
\centering
\fbox{
\includegraphics[trim=0.1cm 0cm 1.4cm 0cm, clip,scale=0.55]{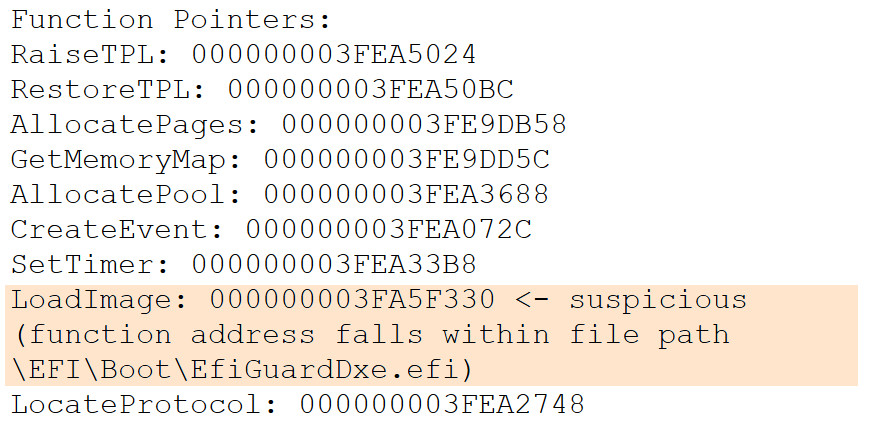}
}
\caption{Boot Services Table analysis showing a suspicious \texttt{LoadImage} function pointer redirected to \texttt{\textbackslash EFI\textbackslash Boot\textbackslash EfiGuardDxe.efi}, indicating potential malicious code execution (full output edited for brevity).}
\label{fig:GluptebaBootServ}
\end{figure}

\begin{figure}[ht]
\centering
\fbox{
\includegraphics[trim=0.1cm 0cm 1.2cm 0cm, clip,scale=0.54]{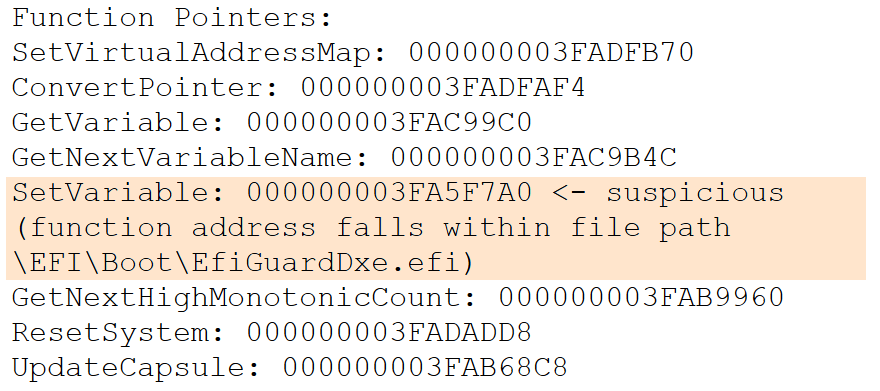}
}
\caption{Runtime Services Table analysis showing a suspicious \texttt{SetVariable} function pointer redirected to \texttt{\textbackslash EFI\textbackslash Boot\textbackslash EfiGuardDxe.efi}, indicating potential malicious activity (full output edited for brevity).}
\label{fig:GluptebaRunServ}
\end{figure}

\vspace{1.5ex}
\noindent\textbf{Detection of the Attack:}  
The module successfully detected malicious modifications introduced by both EfiGuard and its derivative, Glupteba. During the analysis of the Boot Services Table and Runtime Services Table, suspicious function pointer hooks were identified. The detection results are presented in \autoref{fig:GluptebaBootServ} and \autoref{fig:GluptebaRunServ}. For EfiGuard, hooks targeting both \texttt{LoadImage} and \texttt{SetVariable} were flagged, while for Glupteba, only the \texttt{LoadImage} hook was detected. In both cases, the hooks were traced to unexpected memory regions associated with the malicious driver located at \texttt{\textbackslash EFI\textbackslash Boot\textbackslash EfiGuardDxe.efi}, which was used in our attack. 

\subsubsection{CosmicStrand - Hooking via DXE Driver}
\mbox{}\\
\mbox{}\\
\textbf{Execution of the Attack:} Our second attack scenario emulates the hooking methodology of CosmicStrand \cite{CosmicStrand_2022}, based on its documented behavior, as its source code is unavailable. We implemented a malicious DXE driver that modifies randomly selected function pointers in both the Boot Services Table and the Runtime Services Table. The driver, loaded during the DXE phase, accesses the \texttt{EFI\_SYSTEM\_TABLE} to locate both tables. It targets \texttt{AllocatePages}, \texttt{LocateProtocol}, and \texttt{CreateEvent} in the Boot Services Table, and \texttt{GetVariable} and \texttt{SetVariable} in the Runtime Services Table, redirecting their pointers to malicious function handlers within the driver.
The hooking process  is executed with under elevated privileges and disabled interrupts by raising the task priority level (TPL), preventing concurrent access to service tables during modification. 
After replacing the function pointers with malicious versions, the driver recalculates the CRC32 checksum to maintain table integrity. Upon completion, the system resumes the normal boot sequence with the malicious pointers in place.

\begin{figure}[ht]
\centering
\fbox{
\includegraphics[trim=0.12cm 0cm 0.2cm 0cm, clip,scale=0.51]{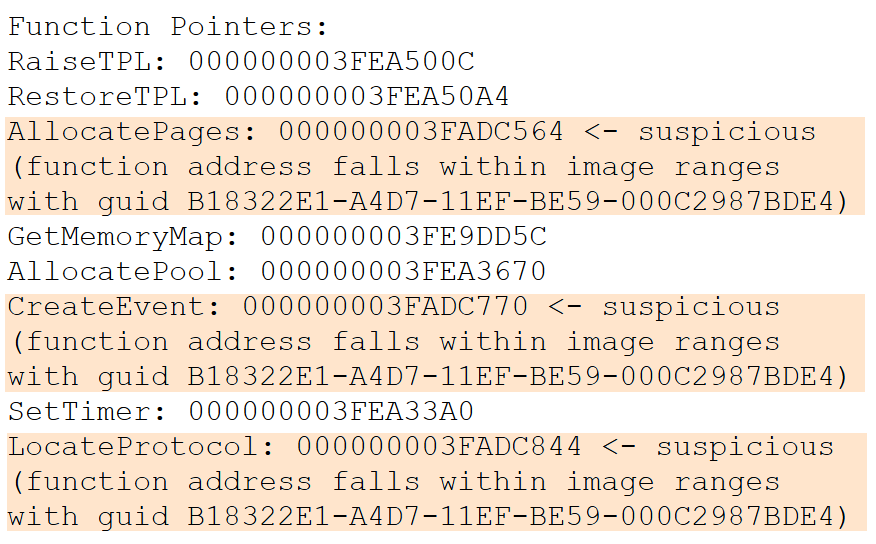}
}
\caption{Boot Services Table analysis showing suspicious function pointers redirected to the memory range of a malicious driver with GUID \texttt{B18322E1-A4D7-11EF-BE59-000C2987BDE4}, indicating a pointer hooking attack (full output edited for brevity).}
\vspace{-0.1cm}
\label{fig:DXEHookBoot}
\end{figure}

\begin{figure}[ht]
\centering
\fbox{
\includegraphics[trim=0.15cm 0cm 0.6cm 0cm, clip,scale=0.53]{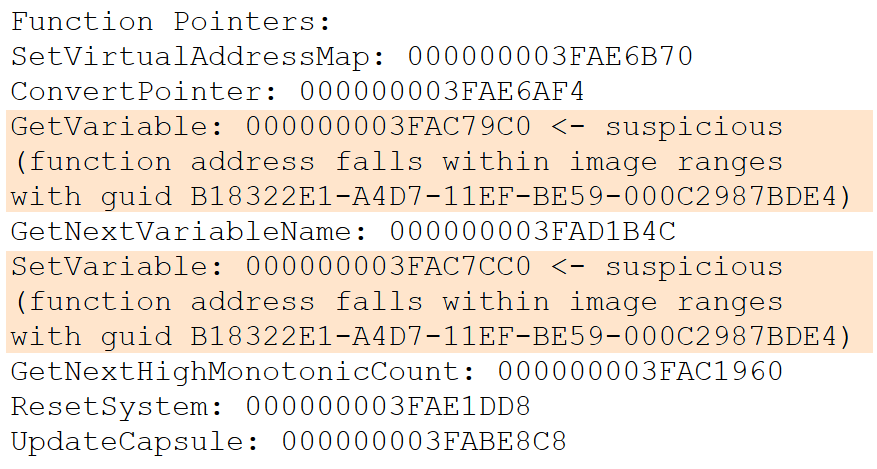}
}
\caption{Runtime Services Table analysis showing suspicious function pointers, redirected to the memory range of a malicious driver with GUID \texttt{B18322E1-A4D7-11EF-BE59-000C2987BDE4}, indicating a pointer hooking attack (full output edited for brevity).}
\label{fig:DXEHookRuntime}
\end{figure}

\vspace{1.5ex}
\noindent\textbf{Detection of the Attack:} Our module detected function pointer modifications in the Boot Services Table and Runtime Services Table. The output of the detection module is presented in \autoref{fig:DXEHookBoot} and \autoref{fig:DXEHookRuntime}. 
Pointers for \texttt{AllocatePages}, \texttt{LocateProtocol}, and \texttt{CreateEvent} in the Boot Services Table and \texttt{GetVariable} and \texttt{SetVariable} in the Runtime Services Table were redirected to unexpected memory regions associated with the injected DXE driver. 
The analysis flagged these anomalies as deviations from the legitimate GUID-defined memory ranges, identifying the malicious driver.

\subsubsection{Thunderstrike - Hooking via OPROM}
\mbox{}\\
\mbox{}\\
\textbf{Execution of the Attack:} This attack, inspired by Thunderstrike \cite{Thunderstrike}, uses a malicious OPROM embedded in a PCI device to hook the \texttt{ProcessFirmwareVolume} function pointer in the DXE Services Table. During the DXE phase, the PCI Bus DXE driver loads the OPROM, which redirects the \texttt{ProcessFirmwareVolume} pointer to malicious code within the OPROM. The attack temporarily elevates the TPL to prevent concurrent access during the modification, ensuring that the hook is applied stealthily. To avoid detection, the CRC32 checksum of the DXE Services Table is recalculated after the pointer modification, maintaining the appearance of integrity.

\begin{figure}[ht]
\centering
\fbox{
\includegraphics[trim=0.15cm 0 0.3cm 0, clip, scale=0.455]{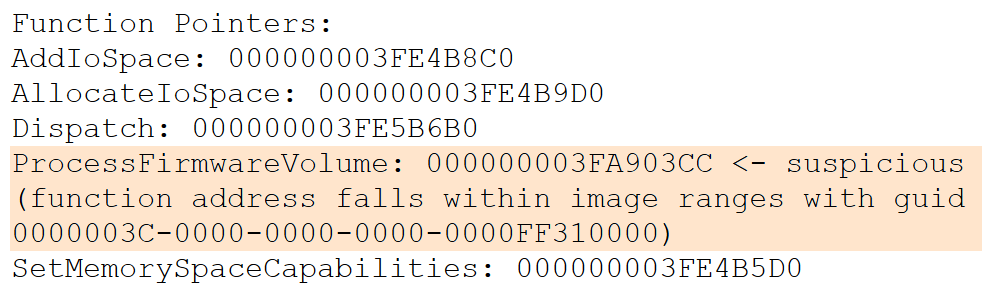}
}
\caption{Detection of a suspicious \texttt{ProcessFirmwareVolume} function pointer redirected to the memory range of a malicious driver with GUID \texttt{0000003C-0000-0000-0000-0000FF310000}, indicating a pointer hooking attack (full output edited for brevity).}
\label{fig:opromDetection}
\end{figure}

\vspace{1.5ex}
\noindent\textbf{Detection of the Attack:} Our detection module detected the pointer redirection to an unexpected memory region within the malicious OPROM. The output of the detection module is presented in \autoref{fig:opromDetection}. By comparing the modified pointer against the GUID-defined memory regions of legitimate drivers, the anomaly was flagged.

\subsection{Inline Hooking Detection}
This module identifies inline hooking by analyzing function prologues for unexpected instruction modifications.

\subsubsection{MoonBounce - Inline Hooking}
\mbox{}\\
\mbox{}\\
\textbf{Execution of the Attack:} In this attack, an inline hooking technique inspired by MoonBounce \cite{MoonBounce_2022} is implemented. We targeted the \texttt{CreateEventEx} function, a routine within the Boot Services Table responsible for event management. Our implementation injects a \texttt{call} instruction to redirect execution to a hard-coded address where our payload resides. This approach enables code execution without altering global service tables, as the modification occurs directly within the function's memory space.
The implementation preserves normal system operation by returning control to the original function after payload execution. While MoonBounce employs an anti-forensics technique by embedding hooks within CORE\_DXE and erasing them post-execution, our implementation maintains the injected code in memory to facilitate the evaluation of detection capabilities, as discussed further in Section~\ref{sec:discussion}.

\vspace{1.5ex}
\noindent\textbf{Detection of the Attack:} Our detection module identified the inline hook through instruction-level examination of \texttt{CreateEventEx}. The disassembly revealed an unexpected \texttt{call} instruction to address \texttt{0x3fadba04}, outside legitimate driver memory regions, as can be seen in Appendix~\ref{app:detection}, in Figure~\ref{fig:InlineHookAttack}. As expected with inline hooking techniques, the Boot Services Table and its reserved fields maintained integrity, yet our instruction-level analysis revealed the memory manipulation that bypassed these traditional integrity checks.

\subsection{UEFI Image Carving}
In addition to hooking, the attacks described above also involved loading of UEFI images from various sources. In our attack implementations, we demonstrated image loading from three distinct locations: the ESP, an OPROM, and a DXE driver embedded directly within the UEFI firmware itself (e.g., stored in SPI flash).
In all cases, the \textit{EFI Image Carving Module} was able to extract the PE files used in each attack from the memory dumps, enabling detailed subsequent analysis of the loaded binaries.

\subsection{False Positive Analysis}
\label{sec:fp_analysis}

To assess practical usability, we evaluated each detection module for false positives across all three test environments: the virtual EDK II setup, the Lenovo ThinkPad, and the System76 Adder WS. This evaluation was conducted on clean systems without malware installed. Table~\ref{tab:fp_results} summarizes the false positive counts for each module (Function pointer hooking, Inline hooking, and Gadget-based detection) across all platforms.

\begin{table}[t]
\centering
\caption{False positive counts across test environments.}
\label{tab:fp_results}
\begin{tabular}{|l|c|c|c|}
\hline
\textbf{Detection Module} & \textbf{Virtual} & \textbf{Lenovo} & \textbf{System76} \\
\hline
\multicolumn{4}{|l|}{\textit{Function Pointer Hooking}} \\
\hline
\quad Before whitelist & 16 & 21 & 16 \\
\quad After whitelist & 0 & 0 & 0 \\
\hline
\multicolumn{4}{|l|}{\textit{Inline Hooking}} \\
\hline
\quad Before GUID filtering & 0 & 1 & 0 \\
\quad After GUID filtering & 0 & 0 & 0 \\
\hline
\multicolumn{4}{|l|}{\textit{Gadget-Based Control-Flow}} \\
\hline
\quad Drivers scanned & 109 & 374 & 76 \\
\quad Baseline thresholds & 15 & 5 & 43 \\
\quad Strict thresholds & 1 & 1 & 1 \\
\hline
\end{tabular}
\end{table}

\subsubsection{Function Pointer Hooking Detection}
As shown in Table~\ref{tab:fp_results}, both the Virtual and System76 platforms produced 16 false positives, all corresponding to the standard EDK II driver hooks. 
The Lenovo platform produced 21 false positives, reflecting five additional vendor-specific hooks.

To address these false positives, this module compares each service table function pointer against a whitelist of known legitimate driver GUIDs. 
The base whitelist contains 16 function-to-GUID mappings, covering six core EDK II drivers:

\begin{itemize}
\item \texttt{PcRtc} for time-related services (\texttt{GetTime}, \texttt{SetTime}, \texttt{GetWakeupTime}, \texttt{SetWakeupTime});
\item \texttt{RuntimeDxe} for address conversion functions (\texttt{SetVirtualAddressMap}, \texttt{ConvertPointer}, \texttt{CalculateCrc32});
\item \texttt{VariableRuntimeDxe} for variable services (\texttt{GetVariable}, \texttt{GetNextVariableName}, \texttt{SetVariable}, \texttt{QueryVariableInfo});
\item \texttt{MonotonicCounterRuntimeDxe} for monotonic counter services (\texttt{GetNextHighMonotonicCount}, \texttt{GetNextMonotonicCount});
\item \texttt{ResetSystemRuntimeDxe} for system reset functionality (\texttt{ResetSystem}); and
\item \texttt{CapsuleRuntimeDxe} for capsule update functions (\texttt{UpdateCapsule}, \texttt{QueryCapsuleCapabilities}).
\end{itemize}

This base whitelist is derived from the UEFI specification and applies to all compliant firmware implementations. 
The whitelist is configurable to accommodate vendor-specific drivers. On the Lenovo ThinkPad platform, we added nine function-to-GUID mappings for three vendor drivers: \texttt{EventCtrl} (\texttt{CreateEvent}, \texttt{SignalEvent}, \texttt{CloseEvent}, \texttt{ExitBootServices}, \texttt{CreateEventEx}), \texttt{LenovoVariableDxe} (\texttt{GetVariable}, \texttt{GetNextVariableName}, \texttt{QueryVariableInfo}), and \texttt{AbtDxe} (\texttt{SetVariable}). 
The whitelist also supports multiple GUIDs per function to accommodate different vendor implementations of the same service.

As shown in Table~\ref{tab:fp_results}, after applying the appropriate whitelist rules, all platforms produced zero false positives.

\subsubsection{Inline Hooking Detection}

This module disassembles the first instructions of each service function and follows control flow transfers (jumps and calls) up to three hops, examining up to 20 instructions per hop. 
A potential hook is flagged when a control transfer targets an address outside the originating driver's memory region.

To reduce false positives from legitimate inter-driver calls, the module compares the GUID of the transfer target against the GUID of the function's originating image. 
Transfers within the same driver (matching GUIDs) are excluded, as they represent internal control flow rather than external hooking. 
This whitelist is configurable; additional trusted GUIDs can be added to accommodate known legitimate cross-driver calls specific to a given platform or vendor.
As shown in Table~\ref{tab:fp_results}, the Virtual and System76 platforms produced zero false positives with the default GUID-based filtering. 
The Lenovo platform produced one false positive from a legitimate cross-driver call.

\subsubsection{Gadget-based Control-Flow Detection}

As described in Section~\ref{sec:rop_detection}, this module extracts gadgets from all carved PE/COFF drivers using Ropper and builds lookup tables mapping gadget offsets and runtime addresses. 
It then scans each driver's memory regions for sequences of 64-bit values that resolve to known gadget addresses. 
Since legitimate driver code and data sections may contain byte patterns that coincidentally match gadget references, the module applies several configurable thresholds to distinguish attack preparations from benign data:

\begin{itemize}
    \item \textbf{Minimum chain length:} The minimum number of consecutive entries required to consider a sequence as a potential chain.
    \item \textbf{Minimum gadget ratio:} The minimum fraction of entries in a candidate sequence that must resolve to valid gadgets.
    \item \textbf{Score threshold:} Each entry is assigned a weight based on its resolution type:
    \begin{itemize}
        \item \textit{gadget} ($+1.0$): value resolves to a known gadget address;
        \item \textit{unknown} ($-0.4$): value appears to be an address but does not resolve to any known gadget;
        \item \textit{data padding} ($0.0$): sentinel values (e.g., null) that attackers commonly use as filler between gadgets; and
        \item \textit{invalid disassembly} ($-0.8$): value resolves to a gadget address, but re-disassembly does not match the expected gadget type.
    \end{itemize}
    The normalized score across all entries must exceed the threshold for the sequence to be flagged.
    \item \textbf{Provider consistency:} When enabled, a minimum fraction of identified gadgets must originate from the same EFI image, reflecting typical ROP chain construction where attackers target gadgets from a single provider module.
\end{itemize}

To evaluate the impact of these thresholds on false positive rates, we conducted a threshold sweep across all three platforms. Table~\ref{tab:gadget_sweep} presents the results. Starting from the baseline configuration shown in Table~\ref{tab:gadget_sweep}, the clean-system false positive counts were 15 (Virtual), 5 (Lenovo), and 43 (System76), totaling 63 flagged regions across 559 scanned drivers (109 Virtual, 374 Lenovo, 76 System76). Progressively tightening the chain criteria reduced false positives, with provider-consistency enforcement producing the largest improvement. The strictest configuration reduced false positives to 1 (Virtual), 1 (Lenovo), and 1 (System76), totaling three flagged regions. 
Analysts can tune these thresholds based on their tolerance for manual verification versus the risk of missing shorter attack chains.

\begin{table}[t]
\centering
\caption{Gadget detection threshold sweep results.}
\label{tab:gadget_sweep}
\scalebox{0.95}{
\begin{tabular}{|l|c|c|c|c|c|c|}
\hline
\textbf{Config} & \textbf{Chain} & \textbf{Gadget} & \textbf{Score} & \textbf{Provider} & \textbf{FPs} & \textbf{Total} \\
 & \textbf{Len.} & \textbf{Ratio} & & \textbf{Cons.} & \textbf{(V/L/S)} & \\
\hline
Baseline & 4 & 0.50 & 0.40 & Off & 15/5/43 & 63 \\
Strict-1 & 5 & 0.60 & 0.50 & Off & 12/3/45 & 60 \\
Strict-2 & 6 & 0.65 & 0.55 & Off & 7/1/31 & 39 \\
Strict-3 & 6 & 0.70 & 0.60 & 0.75 & 2/1/2 & 5 \\
Strict-4 & 8 & 0.75 & 0.65 & 0.80 & 1/1/1 & 3 \\
\hline
\multicolumn{7}{l}{\footnotesize Chain Len. = min\_chain\_length; Gadget Ratio = min\_gadget\_ratio;} \\
\multicolumn{7}{l}{\footnotesize Provider Cons. = provider\_consistency\_ratio (Off = disabled);} \\
\multicolumn{7}{l}{\footnotesize FPs = false positives (Virtual/Lenovo/System76).} \\
\end{tabular}
}
\end{table}

\subsubsection{Summary}

Table~\ref{tab:fp_summary} summarizes the filtering mechanisms and their parameters for each detection module. For the hooking detection modules, false positives represent legitimate drivers or inter-driver calls incorrectly flagged as suspicious. For the gadget-based module, false positives represent memory regions where coincidental byte patterns matched the detection criteria without representing actual attack preparations.

\begin{table}[t]
\centering
\caption{Summary of filtering mechanisms.}
\label{tab:fp_summary}
\scalebox{0.95}{
\begin{tabular}{|l|l|}
\hline
\textbf{Module} & \textbf{Filtering Mechanism} \\
\hline
Function Pointer & Configurable whitelist: 16 base rules (7 EDK II \\
Hooking & drivers); extended with 9 vendor rules on Lenovo \\
\hline
Inline Hooking & Configurable whitelist: automatic GUID-based \\
& filtering; extensible for cross-driver transfers \\
\hline
Gadget-Based & Configurable thresholds (see Table~\ref{tab:gadget_sweep}); \\
Control-Flow & baseline: 63 FPs; strict: 3 FPs \\
\hline
\end{tabular}
}
\end{table}

%% file: Sections/RelatedWork.tex
\section{Related Work}
\subsection{Memory Forensics}
The integration of forensic capabilities at the firmware layer has gained significant traction in recent years \cite{osborne2013memory, vidas2007acquisition, latzo2019universal}. One notable effort is UEberForensIcs, which facilitates OS-level memory acquisition by embedding a DXE driver directly within the UEFI firmware \cite{latzo2021bringing}. Additionally, Intel’s System Management Mode (SMM) has been utilized to obtain memory snapshots with improved integrity and resistance to malware tampering \cite{Reina12}. These capabilities were further extended through the use of PCI network cards in conjunction with SMM, enabling acquisition of both memory and CPU registers to facilitate detailed replication of critical system data \cite{Wang2011}.

Beyond general-purpose memory acquisition, researchers have proposed techniques tailored for specialized environments. For example, memory acquisition methods have been developed for programmable logic controllers (PLCs) \cite{RAIS2021301196,ZUBAIR2022301336,AWAD2023301513} and baseboard management controllers (BMC). One such tool, BMCLeech, enables stealthy memory acquisition from BMCs \cite{latzo2020bmcleech}.
Hardware-based memory approaches have also been explored, particularly for scenarios requiring enhanced security or where firmware access is restricted \cite{BAUER2016S65,  6657268, bulygin2008chipset}. 
Another notable technique involves using a kernel module to directly remap page table entries, granting access to memory regions typically hidden from standard forensic tools \cite{stuttgen2015acquisition}.
\\
Despite these advancements, a significant gap persists in the capability to perform UEFI memory acquisition and forensic analysis during the pre-boot phase. This paper addresses that shortcoming by introducing a specialized memory acquisition and analysis framework designed specifically for the pre-OS environment, offering new capabilities in firmware-level forensics.

\subsection{UEFI Firmware Security Approaches}
To improve the security of UEFI firmware, researchers have developed a wide range of analytical tools and frameworks. Reverse engineering platforms like IDA Pro \cite{IdaPro} and Ghidra \cite{Ghidra}, augmented with plugins like efiXplorer \cite{efixplorer} and efiseek \cite{efiseek}, enable automated disassembly and analysis of UEFI modules. 

Tools like Sentinel One's Brick \cite{Brick} and Binarly's FWHunt \cite{FWHunt} are designed specifically for vulnerability scanning of firmware components, while DXE driver emulators provide environments for controlled testing and behavior inspection. Several security frameworks including Chipsec \cite{chipsec}, UEFITool \cite{uefitool}, and Binwalk \cite{binwalk} focus on UEFI firmware parsing, structure validation, and binary analysis. The field of automated testing has been further advanced through fuzzing techniques. Tools such as AFL \cite{AFL}, TSFFS \cite{TSFFSUEFI}, Efi\_Fuzz \cite{efi_fuzz}, HBFA \cite{HBFA}, and Excite \cite{Excite} have proven effective for uncovering vulnerabilities in UEFI implementations like EDK II \cite{edk2}. 

Academic efforts have further enriched UEFI security with specialized detection methodologies. SPENDER \cite{yin2022finding} and RSFUZZER \cite{yin2023rsfuzzer} were developed for SMM vulnerability detection, while STASE \cite{shafiuzzaman2024stase} combines static analysis with symbolic execution to detect firmware bugs. Additionally, the static analysis framework efiMemGuard \cite{lu2025automated} was introduced to identify memory-related vulnerabilities in UEFI modules.
\\
However, these approaches remain limited to static or emulated analysis and do not provide insight into live firmware memory states during system boot. This lack of dynamic visibility leaves systems vulnerable to runtime-only threats that can evade detection during offline analysis. To bridge this gap, our work introduces a framework for pre-boot memory acquisition and inspection, enabling early detection of malicious modifications during the firmware initialization process.

%% file: Sections/Limitations.tex
\section{Discussion}
\label{sec:discussion}
This section discusses practical considerations, architectural constraints, and current limitations that influence the design and operation of our framework. Some constraints arise from the structure of the UEFI execution environment, while others reflect trade-offs made in the acquisition and analysis components.

\subsection{Constraints of Memory Acquisition}
Memory acquisition during the UEFI phase is shaped by architectural realities rather than tool-specific limitations. Higher-privilege or external acquisition methods are not practical for UEFI forensics. SMM has strict timing limits \cite{6704682} that prevent large memory transfers, and hardware-based approaches such as DMA or cold-boot attacks are unreliable on modern systems due to DMA protections and rapid DRAM decay. OS-level acquisition is unsuitable because the UEFI state is deallocated once \texttt{ExitBootServices} is invoked. As a result, acquiring memory from within the UEFI environment is the only feasible software-based option, even though it shares the same privilege level as potential malware.

\subsection{Implementation Details}
We excluded the dumping of reserved regions by writing zeros to preserve the structure of the dump. This avoids potential instability when accessing reserved memory.

\subsection{Limitations of UEFIMemDump}

The DXE driver version of the dumper must be compiled into the main UEFI firmware. This approach is suitable for virtual environments and for OEM vendors, but it is not applicable on physical systems where the analyst cannot modify the firmware. For such cases, we provide a UEFI application version that can be invoked from the UEFI shell, although this variant operates under different constraints.

The UEFI application runs later in the boot process, since it becomes available only after the UEFI shell is initialized. Earlier acquisition is therefore not possible without firmware integration.

Because acquisition occurs inside the UEFI environment, an attacker who anticipates inspection may attempt to remove traces of their activity or tamper with the dumper, since both run in the same execution context. This introduces a race condition inherent to all software-based acquisition methods, as discussed in \autoref{sec:backgroung:limitations}.

An attacker may also attempt to hide memory regions, for example by modifying \texttt{GetMemoryMap()}. The analysis does not trust any single firmware-reported memory interface and instead derives memory layout from internal UEFI structures. UEFIDumpAnalysis inspects the in-memory service tables directly and resolves each function pointer to its target address. These tables must remain present and internally consistent for the firmware to function, and cross-checking them with the recovered memory layout exposes inconsistencies that reveal concealed regions.

\subsection{Limitations of UEFIDumpAnalysis}
False positives occur across all three detection modules. For function pointer and inline hooking detection, they arise from legitimate drivers that modify service tables or perform cross-driver control-flow transfers; these cases are addressed through configurable GUID-based whitelists that can be extended to accommodate platform or vendor specific drivers and known legitimate cross-driver calls. For gadget-based detection, false positives arise from coincidental byte patterns in code and data sections that match gadget sequences; these cases are addressed through tunable threshold parameters, with tighter thresholds reducing false positives at the cost of potentially missing shorter or sparser attack chains. The false positive rates and the impact of each filtering mechanism are characterized in detail in Section~\ref{sec:fp_analysis}.

%% file: Sections/Conclusions.tex
\section{Conclusions and Future Research}
This paper presents a novel framework for UEFI memory forensics that addresses a gap in below-OS security analysis. The framework consists of two key components: \textit{UEFIMemDump} for memory acquisition and \textit{UEFIDumpAnalysis} for memory analysis. 
Through evaluation with modern UEFI threats such as ThunderStrike, CosmicStrand, and Glupteba, we demonstrated the framework's ability to detect malicious UEFI modifications during the pre-boot phase. Upon acceptance, the framework's source code will be made publicly available. 

We welcome the development of additional detection modules by the research community. While this research focuses on function pointer hooking, inline hooking, and image extraction, the framework was designed to support extensibility. Future modules may focus on entropy-based detection of packed or obfuscated UEFI images, YARA-based scanning to identify known malicious patterns~\cite{lockett2021assessing}, or AI-driven classifiers trained on UEFI memory dumps to uncover anomalous or suspicious code regions. 
Hardware-based memory dumping with an additional layer of robustness could be also addressed in future work.
These additions would further enhance the framework’s forensic capabilities and provide more comprehensive coverage against evolving firmware threats.

%% file: Sections/Acknowledgment.tex
\section*{Acknowledgment}
AI systems were used to assist with text editing, rephrasing, and refinement of code snippets, and all outputs were inspected by the authors to ensure accuracy and originality. 
All technical ideas, experimental design, implementation logic, and analysis were created and validated by the authors.

%% file: Sections/Appendix.tex
\section{UEFI Structures, Attack Techniques, and Detection Output}

\subsection{EFI Boot Services Table}
\label{app:gBS}

Key services of \texttt{gBS} include:

\begin{itemize} [itemsep=0.3em,topsep=2pt,leftmargin=*]

    \item Memory Management Services: Functions such as \texttt{AllocatePages} and \texttt{FreePages} manage physical memory allocation during system initialization;
    
    \item Protocol Management Services: Functions like \texttt{InstallProtocolInterface} and \texttt{LocateProtocol} facilitate interaction with UEFI protocols, which abstract hardware and software services;
    
    \item Event and Timer Services: Functions such as \texttt{CreateEvent}, \texttt{WaitForEvent}, and \texttt{SetTimer} enable event-driven programming and asynchronous operation; and 

    \item Image Services: Functions like \texttt{LoadImage} and \texttt{StartImage} handle the loading and execution of UEFI applications and drivers.

\end{itemize} 

\subsection{EFI Runtime Services Table}
\label{app:gRT}

Key runtime services of \texttt{gRT} include:

\begin{itemize}[itemsep=0.3em,topsep=2pt,leftmargin=*]

    \item Variable Services: Functions such as \texttt{GetVariable}, \texttt{SetVariable}, and \texttt{QueryVariableInfo} manage UEFI variables stored in non-volatile memory for secure configuration and data storage;
    
    \item Time Services: Functions such as \texttt{GetTime} and \texttt{SetTime} handle the system clock and real-time timers; and
    
    \item System Reset Services: The \texttt{ResetSystem} function enables controlled platform resets initiated by the firmware.

\end{itemize}

\subsection{EFI DXE Services Table}
\label{app:gDS}

Key services of \texttt{gDS} include:

\begin{itemize} [itemsep=0.3em,topsep=2pt,leftmargin=*]

    \item Memory Space Management Services: Functions such as \texttt{AddMemorySpace}, \texttt{AllocateMemorySpace}, and \texttt{FreeMemorySpace} dynamically manage physical and virtual memory regions required by DXE drivers;
    
    \item I/O Space Management Services: Functions such as \texttt{AddIoSpace}, \texttt{AllocateIoSpace}, and \texttt{FreeIoSpace} handle the allocation and release of I/O address spaces needed for device communication;

    \item Driver Dispatch Services: Functions like \texttt{Dispatch} and \texttt{Schedule} are used by the DXE Core to load and execute DXE drivers in a controlled and prioritized manner; and 

    \item Firmware Volume Processing Services: The \texttt{ProcessFirmwareVolume} function facilitates the discovery and initialization of firmware volumes, ensuring that all necessary drivers and components are made available during boot.

\end{itemize}

\subsection{Description of UEFI Bootkits Employing Hooking}
\label{app:bootkitsHookingSummary}

\begin{table}[H]
\caption{UEFI bootkits that employ hooking}
\begin{tcolorbox}[
    colback=gray!5,
    colframe=gray!40!black,
    width=\columnwidth,
    arc=5mm,
    boxrule=0.5pt,
    center,
    halign=center
]
\centering
\footnotesize
\begin{tabularx}{\linewidth}{>{\bfseries\raggedright\arraybackslash}p{0.23\linewidth}|>{\raggedright\arraybackslash}X}
\hline
Bootkit & Description \\
\hline
\normalfont MoonBounce \cite{MoonBounce_2022} & MoonBounce is a UEFI bootkit discovered by Kaspersky in 2021. It modifies the CORE\_DXE firmware component in SPI flash, enabling attacks that persist through OS reinstalls and disk replacements. MoonBounce intercepts UEFI functions to load malware to memory, connecting to command-and-control (C2) servers for payload delivery, and is tied to espionage campaigns linked to APT41.\\
\hline
\normalfont CosmicStrand \cite{CosmicStrand_2022} & CosmicStrand is a UEFI rootkit attributed to a Chinese-speaking threat actor. Active since 2016, it infects ASUS and Gigabyte motherboards, modifying the CSMCORE DXE driver to initiate a multi-stage attack. The rootkit sets hooks in the OS loader and Windows kernel, enabling communication with a C2 server for payload delivery. Resilient to OS reinstalls, it primarily targets individuals in China, Vietnam, Iran, and Russia. \\
\hline
\normalfont Glupteba \cite{Glupteba_2024} & Glupteba is a modular malware discovered in the early 2010s, with its UEFI module first seen in 2023. It implants a custom Windows Boot Manager and EfiGuard in the ESP to disable PatchGuard and DSE, enabling persistence. It supports botnet operations, credential theft, and cryptomining. \\
\hline
\normalfont EfiGuard \cite{EfiGuard} & EfiGuard is an open-source UEFI bootkit that patches the Windows Boot Manager, boot loader, and kernel at boot to disable PatchGuard and DSE. It is suspected as the basis for the Glupteba bootkit \cite{Glupteba_2024}. \\
\hline
\normalfont ThunderStrike \cite{Thunderstrike} & Thunderstrike is a PoC UEFI bootkit targeting Apple MacBooks, exploiting vulnerabilities in UEFI firmware to overwrite the SPI flash boot ROM. It hooks the DXE Services Table’s \texttt{ProcessFirmwareVolume} function to intercept and modify firmware updates during recovery mode boots. The attack spreads via writable option ROMs on Thunderbolt devices. \\
\hline
\end{tabularx}
\end{tcolorbox}
\captionsetup{font=footnotesize, width=.9\columnwidth}
\label{tab:hookingBootkitsDesc}
\end{table}

\begin{table}[!ht]
\caption{Hooking in UEFI bootkits.}
\begin{tcolorbox}[
    colback=gray!5,
    colframe=gray!40!black,
    width=\columnwidth,
    arc=5mm,
    boxrule=0.5pt,
    center,
    halign=center
]
\centering
\footnotesize
\begin{tabularx}{\columnwidth}{>{\bfseries\raggedright\arraybackslash}p{0.3\columnwidth}||>{\raggedright\arraybackslash}X}
\hline
Bootkit & Hooking Method \\
\hline\hline
\normalfont MoonBounce \cite{MoonBounce_2022} & Inline hooking of \texttt{gBS->AllocatePool}, \texttt{gBS->CreateEventEx}, and \texttt{gBS->ExitBootServices} \\
\hline
\normalfont CosmicStrand \cite{CosmicStrand_2022} & Function pointer hooking of \texttt{gBS->HandleProtocol} \\
\hline
\normalfont Glupteba \cite{Glupteba_2024} & Function pointer hooking of \texttt{gBS->LoadImage} \\
\hline
\normalfont EfiGuard \cite{EfiGuard} & Function pointer hooking of \texttt{gBS->LoadImage} and \texttt{gRT->SetVariable} \\
\hline
\normalfont ThunderStrike \cite{Thunderstrike} & Function pointer hooking of \texttt{gDS->ProcessFirmwareVolume} \\
\hline
\end{tabularx}
\end{tcolorbox}
\label{tab:hooks}
\end{table}

\subsection{UEFI Structure Parsing}
\label{app:structureparsing}
\begin{figure}[H]
\begin{tcolorbox}[
   colback=gray!5,
   colframe=gray!40!black,
   width=1\columnwidth,
   arc=5mm,
   boxrule=0.5pt,
   center,
   halign=center
]
\includegraphics[scale=0.24]{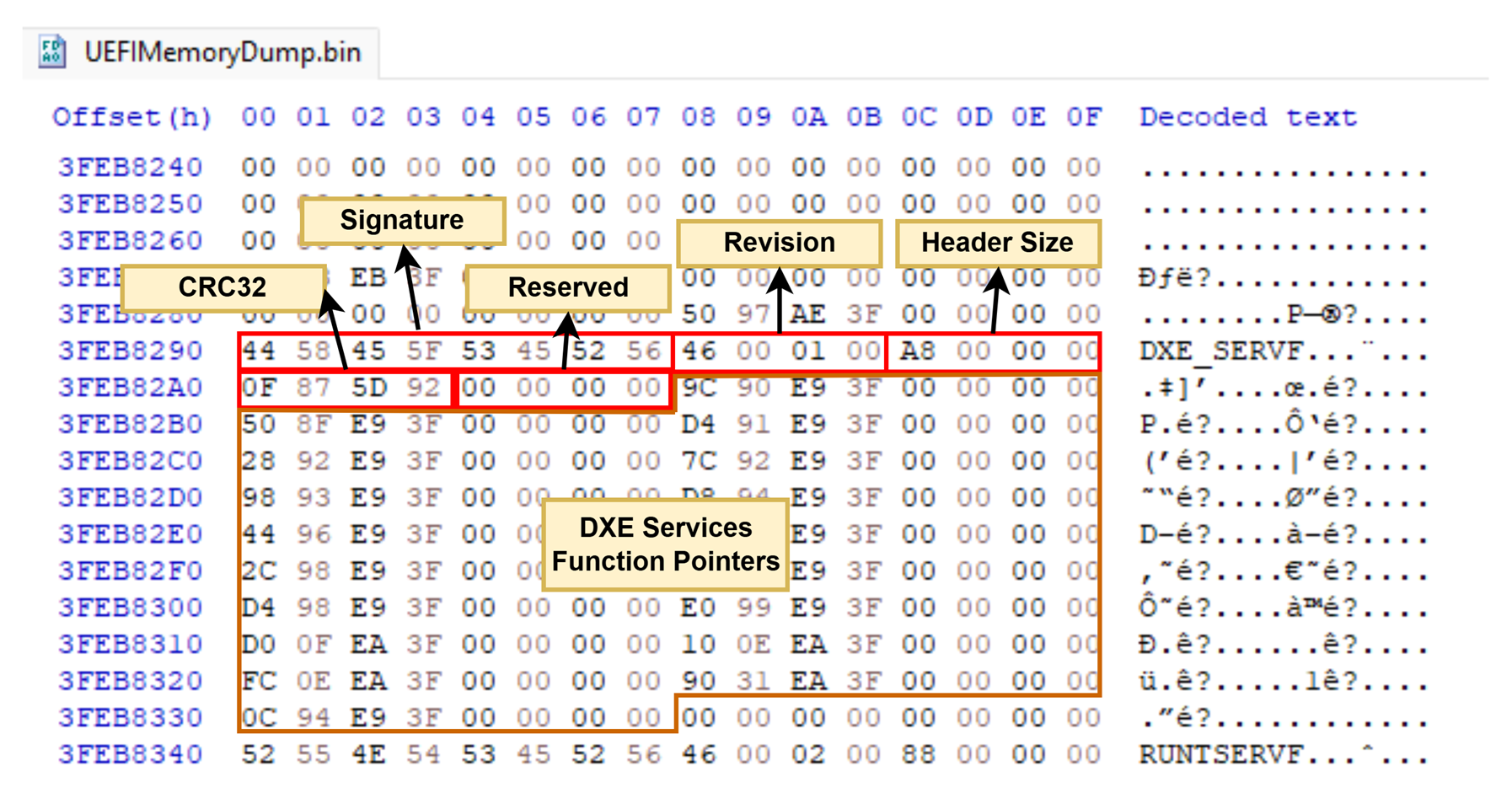}
\end{tcolorbox}
\caption{The in-memory fields of the DXE Services Table, containing metadata and function pointer addresses for the services.}
\label{fig:DxeServ}
\end{figure}

\begin{figure}[ht]
\begin{tcolorbox}[
   colback=gray!5,
   colframe=gray!40!black,
   width=1\columnwidth,
   arc=5mm,
   boxrule=0.5pt,
   center,
   halign=center
]
\includegraphics[scale=0.11]{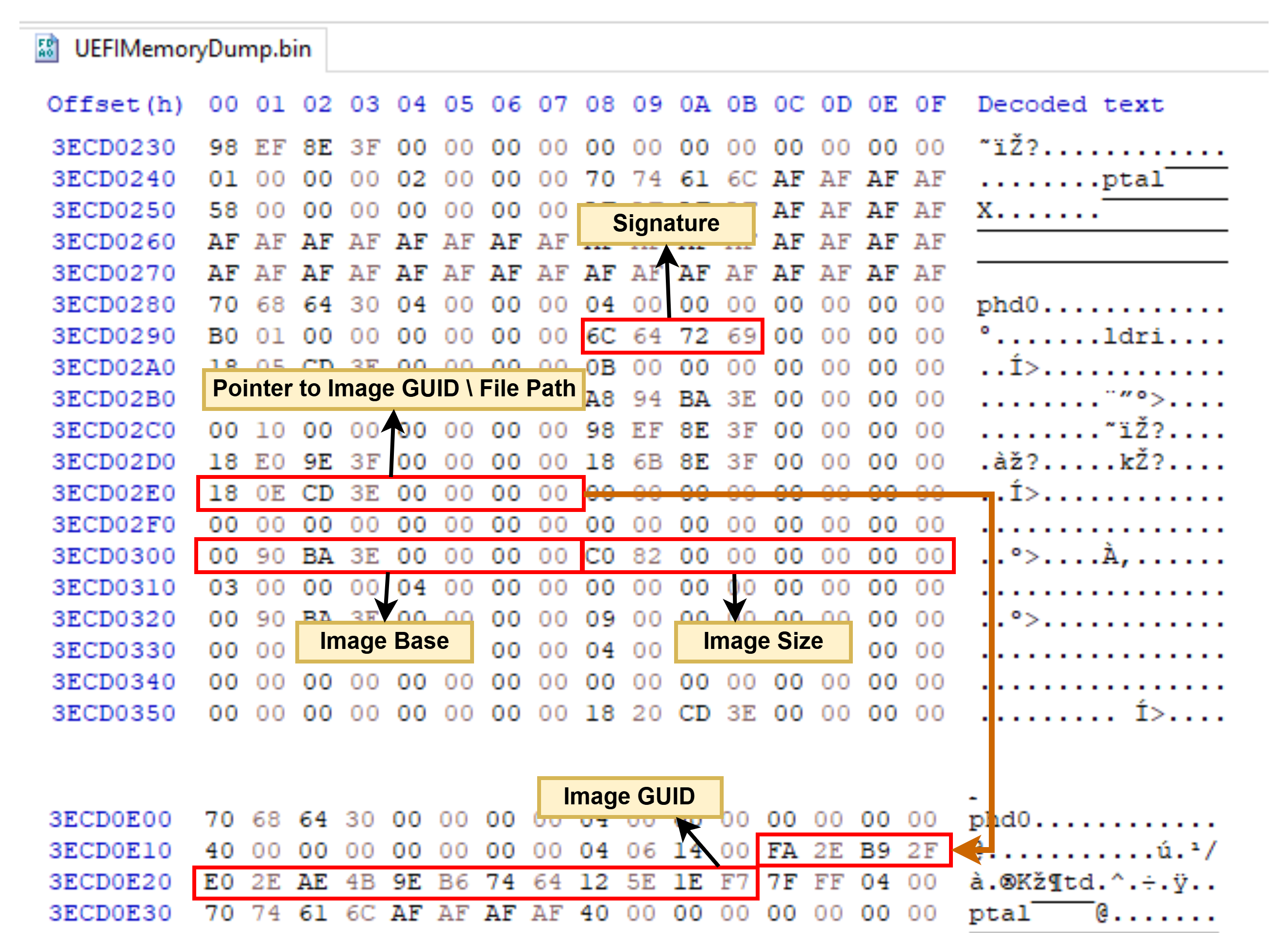}
\end{tcolorbox}
\caption{The in-memory fields of a loaded DXE driver image structure, highlighting the image base, size, and pointer to the image GUID.}
\label{fig:DriverImage}
\end{figure}

\subsection{Detection Results}
\label{app:detection}

\begin{figure}[ht]
\begin{tcolorbox}[
   colback=gray!5,
   colframe=gray!40!black,
   width=\columnwidth,
   arc=5mm,
   boxrule=0.5pt,
   center,
   halign=center
]
\includegraphics[scale=0.48]{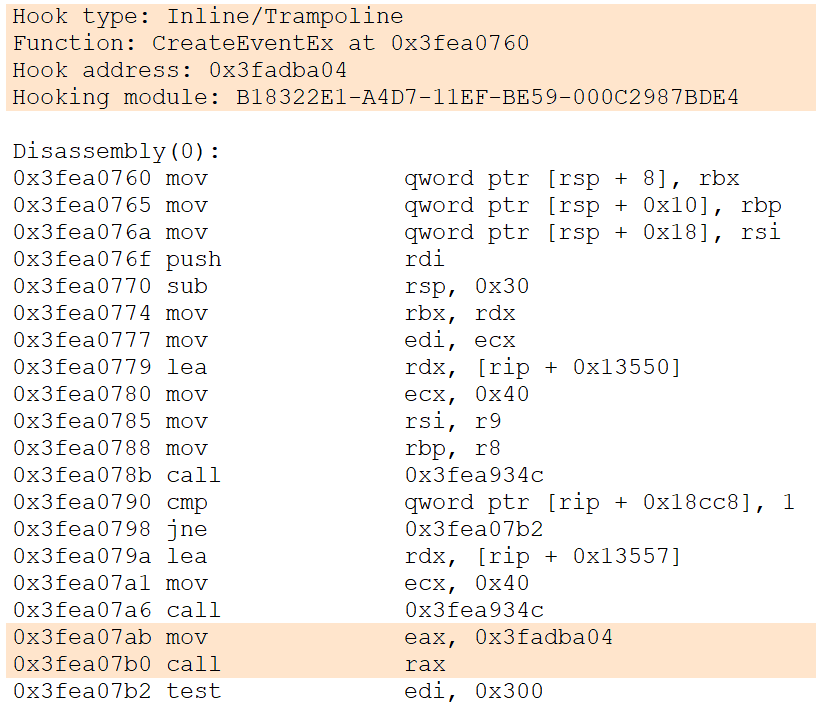}
\end{tcolorbox}
\caption{Disassembly of the \texttt{CreateEventEx} function showing inline hooking redirecting execution to the address \texttt{0x3fadba04} via a \texttt{call} instruction. This modification indicates an in-memory manipulation of the function prologue, allowing execution to be diverted to attacker-controlled code (full output edited for brevity).}
\label{fig:InlineHookAttack}
\end{figure}

\begin{figure}[ht]
\includegraphics[scale=0.35]{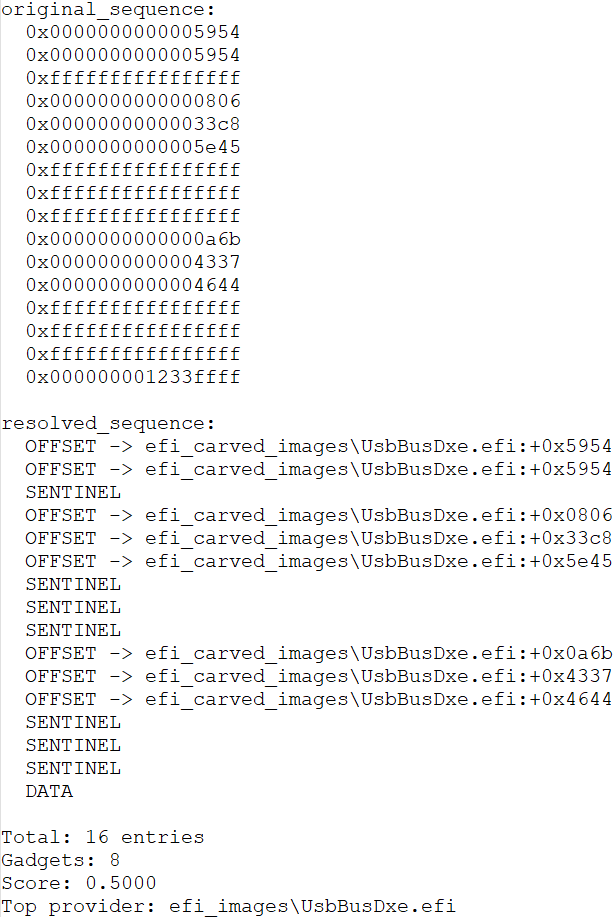}
\captionsetup{font=small, width=0.9\columnwidth}
\caption{Offset-Only Gadget Chain Reconstruction.}
\label{fig:gadgets_offets}
\vspace{-3mm}
\end{figure}

\begin{figure}[ht]
\includegraphics[scale=0.35]{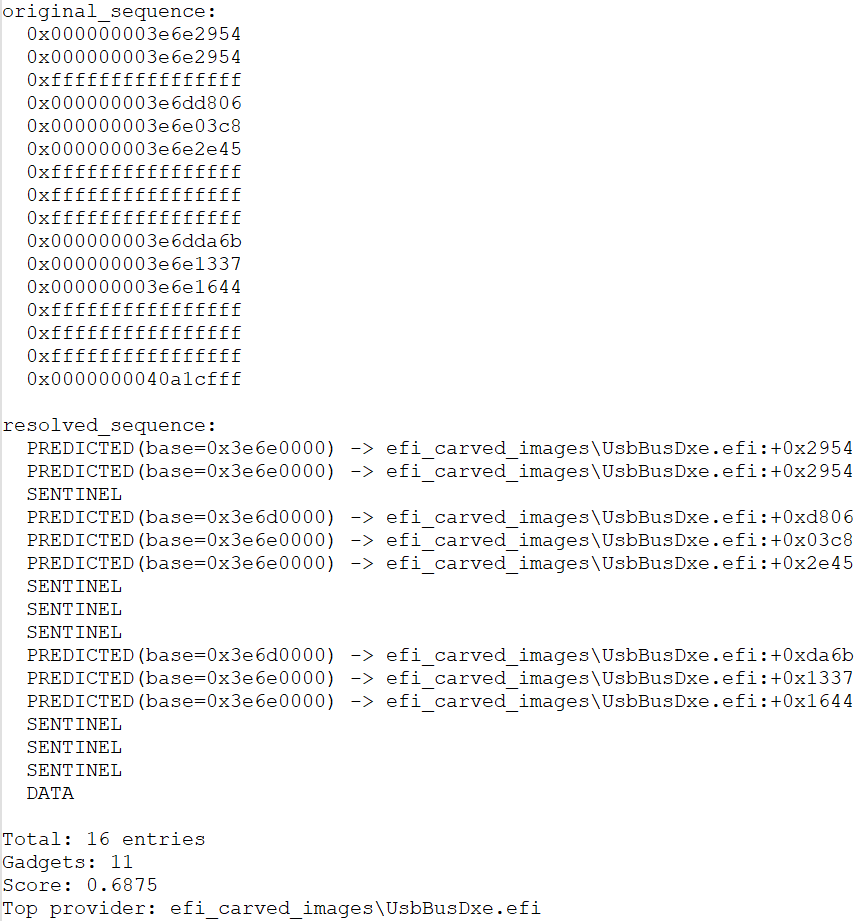}
\captionsetup{font=small, width=0.9\columnwidth}
\caption{Resolved Runtime Gadget Chain Reconstruction.}
\label{fig:gadgets_resolved}
\vspace{-3mm}
\end{figure}